\documentclass[aps,pre,twocolumn,groupedaddress,showpacs,floatfix]{revtex4}

\usepackage{graphicx}% Include figure files
\begin{document}

\title{Reaction-controlled diffusion: Monte Carlo simulations}

\author{Beth A. Reid}
\affiliation{Department of Physics, Virginia Tech, Blacksburg, VA 24061-0435}
%\email{bereid@vt.edu, tauber@vt.edu}

\author{Jason C. Brunson}
\affiliation{Department of Mathematics, Virginia Tech, Blacksburg, 
        VA 24061-0123}

\author{Uwe C. T\"auber}
\affiliation{Department of Physics, Virginia Tech, Blacksburg, VA 24061-0435}

\date{\today}

\begin{abstract}
We study the coupled two-species non-equilibrium reaction-controlled diffusion 
model introduced by Trimper et al. [Phys. Rev. E {\bf 62}, 6071 (2000)] by 
means of detailed Monte Carlo simulations in one and two dimensions.
Particles of type $A$ may independently hop to an adjacent lattice site
provided it is occupied by at least one $B$ particle.
The $B$ particle species undergoes diffusion-limited reactions.
In an active state with nonzero, essentially homogeneous $B$ particle
saturation density, the $A$ species displays normal diffusion.
In an inactive, absorbing phase with exponentially decaying $B$ density, the
$A$ particles become localized.
In situations with algebraic decay $\rho_B(t) \sim t^{- \alpha_B}$, as occuring
either at a non-equilibrium continuous phase transition separating active and
absorbing states, or in a power-law inactive phase, the $A$ particles propagate
subdiffusively with mean-square displacement
$\langle {\vec x}(t)^2_A \rangle \sim t^{1 - \alpha_A}$.
We find that within the accuracy of our simulation data,
$\alpha_A \approx \alpha_B$ as predicted by a simple mean-field approach.
This remains true even in the presence of strong spatio-temporal fluctuations
of the $B$ density.
However, in contrast with the mean-field results, our data yield a distinctly
non-Gaussian $A$ particle displacement distribution $n_A({\vec x},t)$ that
obeys dynamic scaling and looks remarkably similar for the different processes
investigated here.
Fluctuations of effective diffusion rates cause a marked enhancement of
$n_A({\vec x},t)$ at low displacements $|{\vec x}|$, indicating a considerable
fraction of practically localized $A$ particles, as well as at large traversed
distances.
\end{abstract}


% 05.40.-a  Fluctuation phenomena, random processes, noise, and Brownian motion
% 05.40.Fb  Random walks and Levy flights
% 64.60.Ht  Dynamic critical phenomena
% 82.20.-w  Chemical kinetics and dynamics

\maketitle

\section{Introduction}
\label{intro}

The goal of statistical mechanics is to understand the relationship between
microscopic and macroscopic dynamics in systems consisting of a large number
of degrees of freedom.
One classical success of the equilibrium formalism is the prediction of
universal phase transition behavior:
independent of the microscopic details of their interactions, systems with
identical {\em overall} features, governed by their symmetries, spatial
dimension $d$, and perhaps large-scale interaction properties, display very
similar phase diagrams.
Moreover, their critical points are characterized by the same small set of
independent scaling exponents.
Thus physical systems with very complicated interactions can often be
adequately described by considerably simplified models, which in turn form
the basis of simulation studies and numerical analysis.

Here we investigate a simple coupled reaction-diffusion system, which however 
leads to remarkably rich features.
More specifically, the spatio-temporal fractal structures emerging at a 
{\em non-equilibrium} critical point of a reacting species $B$ impose 
non-trivial scaling behavior onto the propagation of passive random walkers 
$A$, whose propagation is however limited to sites occupied by at least one 
$B$ particle.
One may envision this system to model virus (represented by the $A$ particles)
propagation in a carrier $B$ population that is set close to its extinction 
threshold; the virus remains dormant when there are no $B$ organisms present.
Below, we shall encounter and characterize the ensuing scaling laws by means
of Monte Carlo simulations, and compare our numerical results with the
predictions of a mean-field approximation. 

In non-equilibrium systems, the detailed balance conditions are violated;
i.e., the probability of at least one closed loop of transitions between
microscopic configurations depends upon the direction the loop is traversed.
This is the case even in stationary states in open systems, through which a
steady particle or energy current from the outside is maintained.
Outside physics non-equilibrium models may describe, for example, population
dynamics, chemical catalysis, and financial markets.
Yet reassuringly, universal behavior has also been found to persist for
non-equilibrium models that display phase transitions between different
stationary states.

Prominent examples are continuous transitions between active and
inactive/absorbing states in diffusion-limited `chemical' reactions
\cite{hinrichsen00}.
The class of models we will be studying involves competing annihilation and 
offspring reactions of a single species $B$, performing unbiased random walks 
on a $d$-dimensional hypercubic lattice:
\begin{eqnarray}
  B & \stackrel{\lambda}{\rightarrow} & \emptyset \nonumber \\
  B & \stackrel{\sigma}{\rightarrow} & (m + 1) \, B \label{rxs} \\
  n \, B & \stackrel{\mu}{\rightarrow} & \emptyset \ , \nonumber
\end{eqnarray}
with integers $m \geq 1$, $n > 1$.
For large branching rate $\sigma$, the system is in an active state with a
non-zero and essentially homogeneous particle density.
In contrast, when the decay processes with rates $\lambda$ and $\mu$ dominate,
the $B$ particle density reaches zero, and the dynamics ceases entirely in
this inactive, {\em absorbing} state.
By appropriately tuning the reaction rates a continuous phase transition
between these two stationary states can be observed \cite{fnote3}.

Generically, such transitions fall into the {\em directed percolation} (DP)
universality class \cite{janssen81} with upper critical dimension $d_c = 4$.
The standard example is represented by the Gribov process $B  \to \emptyset$,
$B \rightleftharpoons 2 B$.
Equivalently, one may use the scheme (\ref{rxs}) with $m = 1$ and $n = 2$.
Directed percolation was initally devised to characterize the transition from
finite to infinite-sized clusters in directed media (such as a porous rock in
a gravitational field) \cite{kinzel83}.
Other applications include certain models of catalytic reactions, interface
growth, turbulence, and the spread of epidemics \cite{hinrichsen0b}.
Experimental evidence for DP critical behavior was recently observed in
spatio-temporal intermittency in ferrofluidic spikes \cite{rupp03}.

For active to absorbing state phase transitions in single-species reactions
that include first-order processes, in the absence of memory effects and
quenched disorder, the so-called {\em parity conserving} (PC) universality
class appears to represent the only scenario for non-DP critical behavior
\cite{hinrichsen00}.
In the above reaction scheme (\ref{rxs}), PC scaling is observed for branching
and annihilating random walks with $\lambda = 0$, $n = 2$, and {\em even}
offspring number $m$.
In that situation, reactions either create or annihilate an even number of
particles.
Thus the number of B particles remains either even or odd throughout the
system's temporal evolution.
Indeed, the distinct non-trivial scaling exponents of the PC universality
class, albeit limited essentially to $d = 1$, can be attributed to this special
symmetry of local particle number parity conservation.
Moreover, for $d \leq d_c' \approx 4/3$, fluctuations cause the emergence of a
power-law inactive phase, characterized by the algebraic decay laws of
diffusion-limited pair annihilation $2 \, B \to \emptyset$
($\lambda = \sigma = 0$, $n = 2$) \cite{cardy96}.

When $\lambda > 0$ or $m$ is odd (for $n = 2$), however, parity conservation is
destroyed.
The case $\lambda > 0$ immediately yields a transition in the DP universality
class with $d_c = 4$.
Yet for odd $m$, even if $\lambda = 0$ initially, fluctuations {\em generate}
sufficiently strong decay processes $B \to \emptyset$ in $d \leq 2$ dimensions
to produce a transition to an inactive phase with DP critical exponents.
For $\lambda = 0$ and $d > 2$, one encounters only an active phase for any
$\sigma > 0$, as predicted by the mean-field rate equation \cite{cardy96}.

\begin{figure}
\includegraphics*[scale=0.7,angle=0]{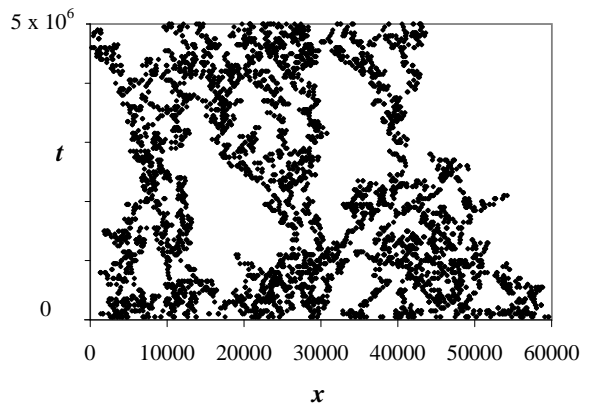}
\caption{\label{fig:clus} Space-time plot for the one-dimensional reactions
  $B \to 3 \, B$, $2 \, B \to \emptyset$ at the active/absorbing critical
  point (PC universality class).}
\end{figure}

At the non-equilibrium continuous phase transition, the reacting particles
form spatio-temporal fractal structures characterized by algebraic decay of
the correlation functions (one example is depicted in Fig.~\ref{fig:clus}).
In Ref.~\cite{trimper00} it was suggested to employ these {\em dynamic}
fractals of reacting agents $B$ as backbones for nearest-neighbor hopping
processes of another, otherwise passive particle species $A$.
The $A$ particles are then allowed to move to a nearest neighbor site only if
that site is occupied by at least one $B$ particle.
In an active state of the $B$ system, with a largely homogeneous particle
distribution, the $A$ particles follow Fick's normal diffusive propagation
law, $\langle {\vec x}(t)^2_A \rangle = 2 \, D \, t$, with diffusion constant
$D \sim a_0^2 \, / \tau_0$, where $a_0$ denotes the lattice constant of the
hypercubic lattice, and $\tau_0^{-1}$ the microscopic hopping rate.
On the contrary, in a DP-type inactive phase with an exponentially decaying
$B$ particle density, the $A$ species will become localized, i.e.,
$\langle {\vec x}(t)^2_A \rangle \to {\rm const}$ as $t \to \infty$.
In precisely this sense the inactive to active state transition of the $B$ 
system thus induces a {\em localization transition} for the $A$ particles.
At the transition itself, as well as in the PC-inactive phase, the $B$ density
decays algebraically,
\begin{equation}
  \rho_B(t) \sim t^{- \alpha_B} \ , \label{bds}
\end{equation}
with $0 < \alpha_B \leq 1$.
Correspondingly, the $A$ species propagates {\em subdiffusively},
\begin{equation}
  \langle {\vec x}(t)^2_A \rangle \sim t^{1 - \alpha_A} \ , \label{amd}
\end{equation}
where again $0 < \alpha_A \leq 1$.
In fact, a simple mean-field approach suggests $\alpha_A = \alpha_B$
\cite{trimper00}.
Our goal here is to further elucidate the scaling laws for the ensuing
anomalous $A$ particle diffusion through Monte Carlo simulations in one and two
dimensions.
We shall also numerically determine the full time-dependent probability
distribution for the $A$ species displacements and compare it with the Gaussian
distribution predicted by mean-field theory \cite{trimper00}.

Before we proceed, we note that our model is related to, but quite distinct
from studies of anomalous diffusion on {\em static} fractal structures.
These have been used to describe a variety of physical phenomena, such as
percolation through porous or fractured media and electron-hole recombination
in amorphous semiconductors \cite{bunde92}.
Anomalous diffusion may also ensue for particle transport in a random medium
with quenched disorder, provided the obstacles are sufficiently strong to
effectively reduce the number of diffusive paths on large length scales
\cite{havlin87}.
We emphasize again that in the system under investigation here the fractal 
structure evolves {\em dynamically}, which provides an alternative mechanism 
for subdiffusive propagation: 
When Eq.~(\ref{bds}) holds, the number of available paths decreases with time.
Yet notice that $A$ particles that have become localized on an isolated $B$
cluster for some time may still become linked to a large connected region of
available sites later.

\begin{table*}
\caption{Critical exponents for the parity conserving (PC) and directed
  percolation (DP) universality classes of active to absorbing phase
  transitions, as determined from Monte Carlo simulations in one and two
  dimensions [1].
  Here, $r$ denotes the deviation of a relevant control parameter from the
  critical point.
  For directed percolation, the first-order results from the $\epsilon$
  expansion near $d_c = 4$ dimensions are given as well.}
\label{exptb}
\begin{ruledtabular}
\begin{tabular}{ccccc}
Critical exponents & PC, $d = 1$ & DP, $d = 1$ & DP, $d = 2$ 
& DP, $d = 4 - \epsilon$ \\ \hline
$\rho_s \sim |r|^\beta$ & $\beta \approx 0.92$ & $\beta \approx 0.276$
& $\beta \approx 0.584$ & $\beta = 1 - \epsilon/6 + O(\epsilon^2)$ \\
$\xi \sim |r|^{-\nu}$ & $\nu \approx 1.84$ & $\nu \approx 1.097$
& $\nu \approx 0.734$ & $\nu = 1/2 + \epsilon/16 + O(\epsilon^2)$ \\
$t_c \sim \xi^z \sim |r|^{-z \nu}$ & $z \approx 1.75$ & $z \approx 1.581$
& $z \approx 1.764$ & $z = 2 - \epsilon/12 + O(\epsilon^2)$ \\
$\rho_c(t) \sim t^{-\alpha}$ & $\alpha \approx 0.285$ & $\alpha \approx 0.159$
& $\alpha \approx 0.451$ & $\alpha = 1 - \epsilon/4 + O(\epsilon^2)$ \\
\end{tabular}
\end{ruledtabular}
\end{table*}

In the subsequent Sec.~\ref{theor} we briefly review the theoretical
considerations of Ref.~\cite{trimper00}, and list the central results from
the mean-field approach for the $A$ species propagation on the dynamic $B$
fractal.
In Sec.~\ref{simul} we give an overview of the Monte Carlo simulation methods 
employed in our study.
Next, in Sec.~\ref{nihil} we first present our results for the anomalous $A$
diffusion as induced by pure annihilation kinetics of the $B$ species
($\sigma = 0$).
Section~\ref{trans} is devoted to the central issue in our investigation,
namely the subdiffusive behavior of the $A$ particles at the localization
transition caused by the active to inactive/absorbing phase transition of the
reacting agents $B$.
We summarize our results in Sec.~\ref{concl}, add concluding remarks, and
point out a few open problems.

\section{Theoretical predictions}
\label{theor}

For the specific combination of $B$ particle reactions studied here, the
asymptotic scaling behavior is well understood.
In particular, we shall consider both systems with pure annihilation kinetics
($n \, B \rightarrow \emptyset$), and models with competing annihilation and
offspring reactions, as listed in (\ref{rxs}), in the vicinity of their
non-equilibrium phase transition from active to inactive, absorbing states.
These phase transitions either fall into the directed percolation (DP) or
parity conserving (PC) universality classes, both of which have been
previously investigated in detail \cite{hinrichsen00}.
Specifically, the exponents $\alpha_B$ characterizing the long-time decay of
the particle density are well-established through simulations (see
Tab.~\ref{exptb}).

Let us first consider pure annihilation kinetics with reaction rate $\mu$.
The corresponding mean-field rate equation for the $B$ particle density reads
\begin{equation}
  \partial_t \, \rho_B(t) = - n \, \mu \, \rho_B(t)^n \ . \label{mfb}
\end{equation}
For $n > 1$ this yields
\begin{equation}
  \rho_B(t) = \frac{\rho_B(0)}{\left( 1 + t / \tau_n \right)^{1/(n-1)}} \ ,
  \quad \tau_n = \frac{\rho_B(0)^{1-n}}{n(n-1) \, \mu} \ , \label{mas}
\end{equation}
whence at sufficiently large times $\rho_B(t) \sim (\mu t)^{-1/(n-1)}$,
independent of the initial density $\rho_B(0)$.
For n = 1, i.e., spontaneous death with rate $\lambda$, one naturally finds
exponential decay,
\begin{equation}
  \rho_B(t) = \rho_B(0) \ e^{- \lambda t} \ . \label{mrd}
\end{equation}

We expect these results to be valid (at least qualitatively) in dimensions
above an upper critical dimension $d_c$, which can be determined through
straightforward dimensional analysis:
since the $B$ particles undergo ordinary diffusion, we expect $[t] = [x]^2$,
and of course in $d$ dimensions, $[\rho_B] = [x]^{-d}$.
Then by Eq.~(\ref{mfb}), $[\mu] = [x]^{d (n-1) - 2}$.
Thus $\mu$ is dimensionless (marginal in the RG sense) at $d_c(n) = 2/(n-1)$.
The mean-field description (\ref{mfb}) fails in lower dimensions where there
is a non-negligible probability that a particle will retrace part of its
trajectory.
For $n = 2, 3$ anti-correlations between surviving particles are induced in
dimensions $d \leq 2$ and $d = 1$, respectively, since many of the nearby
particles along a specific agent's trajectory are annihilated in the first 
trace.
The annihilation processes then become diffusion- rather than reaction-limited.
For the pair process, the particles need to traverse a distance
$\ell(t) \sim (D t)^{1/2}$ before they can meet and annihilate, where $D$
denotes the $B$ particle diffusion constant.
Hence the particle density should scale as 
$n(t) \sim \ell(t)^{-d} \sim (D t)^{-d/2}$.
Indeed, a renormalization group analysis predicts for $2 B \to \emptyset$
(as well as for pair coagulation $2 B \to B$) \cite{lee94}
\begin{eqnarray}
  &\rho_B(t) \sim (D t)^{-d/2} \quad &{\rm for} \ d < 2 \ , \label{pad} \\
  &\rho_B(t) \sim (D t)^{-1} \, \ln D t \quad &{\rm at} \ d_c(2) = 2 \ ,
\label{pa2}
\end{eqnarray}
in agreement with exact solutions in $d=1$.
Thus particles survive considerably longer than Eq.~(\ref{mas}) would suggest.
For triplet annihilation ($3 B \to \emptyset$), in one dimension there remain
mere logarithmic corrections to the mean-field result \cite{lee94},
\begin{equation}
  \rho_B(t) \sim \left( \frac{\ln D t}{D t} \right)^{1/2} \quad {\rm at} \
  d_c(3) = 1 \ . \label{ta1}
\end{equation}
The higher-order ($n \geq 4$) annihilation processes should all be aptly
described by the mean-field power laws (\ref{mas}).

Exact results for the critical behavior of the DP and PC universality classes
cannot be derived analytically, but the scaling exponents have been measured
quite accurately by means of computer simulations \cite{hinrichsen00}, see
Tab.~\ref{exptb}.
The universal properties of directed percolation can be represented through
Reggeon field theory \cite{cardy80}, which allows a systematic perturbational
calculation of the critical exponents in an $\epsilon$ expansion near its upper
critical dimension $d_c = 4$.
The one-loop fluctuation corrections to their mean-field values are listed in
Tab.~\ref{exptb} as well.
At least for DP, the scaling relation $\beta = z \, \nu \, \alpha$ holds.
A similarly reliable analytic computation of the PC critical exponents has as
yet not been achieved, owing to the absence of a corresponding mean-field 
theory (see Ref.~\cite{cardy96} for further details).

In Ref.~\cite{trimper00}, the reaction-controlled diffusion model was defined
as follows.
Otherwise passive agents $A$ perform independent random walks to those sites
that are occupied by at least one $B$ particle; the $B$ species is subject to
diffusion-limited reactions of the above type.
In our simulations, we have assumed the $A$ hopping rate $\tau_0^{-1}$ to be
independent of the number of $B$ particles at adjacent sites.
This contrasts the model of Ref.~\cite{trimper00}, where the $A$ hopping
probability to a given site was taken to be proportional to the number of
$B$ particles on that site.
Yet as we are mostly interested in the asymptotic behavior at low $B$ 
densities, this distinction should be largely irrelevant.
Moreover, in the majority of systems studied here the $B$ pair annihilation
reaction was set to occur with probability $1$, which eliminates multiple site
occupations.

To be specific, consider $B$ particles undergoing the Gribov reactions
$B \to \emptyset$, $B \rightleftharpoons 2 B$, with an ensuing critical
point in the DP universality class.
The effective theory near the phase transition then becomes equivalent to a
Langevin equation for a fluctuating field $b({\vec x},t)$
\cite{cardy80, janssen81, fnote1}:
\begin{equation}
  \partial_t b = D \, (\nabla^2 - r) \, b - 2 \mu \, b^2 + \eta \ . \label{dpl}
\end{equation}
With $\langle \eta({\vec x},t) \rangle = 0$, the ensemble average of $b$ over
noise realizations yields the mean $B$ particle density,
$\langle b({\vec x},t) \rangle = \rho_B(t)$.
For the correlator of the stochastic noise one finds
\begin{equation}
  \langle \eta({\vec x},t) \, \eta({\vec x'},t') \rangle = 2 \, \sigma \,
  b(\vec x,t) \, \delta({\vec x} - {\vec x'}) \, \delta(t-t') \ , \label{dpn}
\end{equation}
which is to be understood as the prescription to always factor in the local
particle density when noise averages are taken.
According to Eq.~(\ref{dpn}), all fluctuations vanish in the absorbing state,
as they should.
As before, $\lambda$, $\sigma$, and $\mu$ represent the $B$ particle decay,
branching, and coagulation rates, respectively.
The control parameter $r$ denotes the deviation from the critical point, e.g.,
$r = (\lambda - \sigma) / D$ in the mean-field approximation.

With the model definition in Ref.~\cite{trimper00}, the effective diffusivity
of the agents $A$ becomes proportional to the local $B$ density.
In fact, starting from the classical master equation, one can derive the
following continuum stochastic equation of motion for a coarse-grained field
$a({\vec x},t)$ that describes the $A$ species \cite{trimper00}
\begin{equation}
  \partial_t a = {\widetilde D} \, (\nabla^2 a) \, b
  - {\widetilde D} \, a \, (\nabla^2 b) + \zeta \ , \label{lda}
\end{equation}
with noise correlations
\begin{eqnarray}
  &&\!\!\!\!\! \langle \zeta({\vec x},t) \, \zeta({\vec x'},t') \rangle = 0 \ ,
  \label{noi} \\
  &&\!\!\!\!\! \langle \zeta({\vec x},t) \, \eta({\vec x'},t') \rangle =
  {\widetilde D} \, [\nabla^2 a({\vec x},t)] \, b({\vec x},t) \,
  \delta({\vec x} - {\vec x'}) \, \delta(t-t') \nonumber \\
  &&\qquad\qquad\quad\ - {\widetilde D} \, a(\vec x,t) \, \nabla^2
  [b({\vec x},t) \, \delta({\vec x} - {\vec x'}) \, \delta(t-t')] \ . \nonumber
\end{eqnarray}
The fluctuations of the $b$ field thus influence the $A$ diffusion in a
non-trivial manner.

Certainly outside the critical regime, well inside either the active or
inactive phases, which for DP are both characterized by exponentially decaying
correlations in space and time, one may apply a mean-field type of
approximation.
To this end, we consider the $B$ particle density to be spatially homogeneous,
and neglect the noise cross-correlations in Eq.~(\ref{noi}).
Upon rescaling, the equation of motion (\ref{lda}) then reduces to a mere
diffusion equation \cite{trimper00}
\begin{equation}
  \partial_t \, n_A({\vec x},t) = 
  {\widetilde D} \, \rho_B(t) \, \nabla^2 \, n_A({\vec x},t)
\label{dif}
\end{equation}
for the probability $n_A$ of finding a particle $A$ at point ${\vec x}$ at
time $t$, with time-dependent effective diffusivity
${\widetilde D} \, \rho_{B}(t)$.
We may interpret this result as follows.
In the original model, the hopping rate to an adjacent site is proportional to
the number of $B$ particles on that site.
At sufficiently low densities that multiple $B$ particle occupation of a given
site can be neglected, the average $B$ density represents the fraction of
lattice sites available for the $A$ particles to hop to.
Thus we expect the $A$ species diffusion rate to be approximately proportional
to the global $B$ particle density.
For this assumption to be accurate, the $A$ particle distribution must also be
assumed to be at least roughly uniform.
However, local fluctuations of the $B$ density may induce some clustering for
many $A$ particles as well, though certainly to a lesser degree since in 
regions with small disjoint $B$ clusters, the $A$ particles become localized.
Any deviations from the uniform effective diffusion coefficient caused by an
inhomogeneous $B$ particle distribution would be diminished by this
simultaneous clustering.
We will discuss these effects in more detail below as we measure deviations 
from the mean-field behavior.

Under the above mean-field assumption, we may employ the diffusion equation
(\ref{dif}) to determine how the probability distribution $n_{A}(\vec x,t)$
evolves in time for a system of independent $A$ particles that all start
initially at a particular location ${\vec x} = 0$, i.e.,
$n_{A}({\vec x},0) = \delta({\vec x})$.
Even with a time-dependent diffusion coefficient, Eq.~(\ref{dif}) is readily
solved via spatial Fourier transformation, resulting in a Gaussian:
\begin{equation}
  n_{A}({\vec x},t) = \frac{1}{\left[ 4 \pi \, D'(t) \right]^{d/2}} \
  \exp \left( - \frac{{\vec x}^2}{4 \, D'(t)} \right) \ . \label{sol}
\end{equation}
But the expression $D t$ in Fick's law for standard diffusion becomes
replaced with an integral over the evolving $B$ density,
\begin{equation}
  D'(t) = {\widetilde D} \int_0^t \rho_B(t') \, dt' \ . \label{dpt}
\end{equation}
Naturally, the odd moments of the distribution (\ref{sol}) vanish, while
$\langle {\vec x}(t)^k_A \rangle = \langle |{\vec x}(t)_A|^k \rangle > 0$ for
$k$ even.
We then compute
\begin{eqnarray}
  &&\langle |{\vec x}(t)_A|^k \rangle =
  \int |{\vec x}|^k \, n_A({\vec x},t) \, d^dx \nonumber \\
  &&\qquad\qquad\quad\! = \frac{[4 \, D'(t)]^{k/2}}{\Gamma(d/2)} \
  \Gamma\!\left(\frac{k+d}{2} \right) \ . \label{mgm}
\end{eqnarray}
For $k = 2$ in particular, this reduces to
\begin{equation}
  \langle {\vec x}(t)^2_A \rangle = 2 d \, D'(t) \ . \label{m2s}
\end{equation}

For a constant $B$ particle density, e.g., the saturation value $\rho_s$ in an
active phase, we recover ordinary diffusion with effective diffusivity
$D = {\widetilde D} \rho_s$.
In an inactive phase with exponential density decay (\ref{mrd}), we find 
instead
\begin{equation}
  \langle {\vec x}(t)^2_A \rangle =
  \frac{2 d \, {\widetilde D} \, \rho_B(0)}{\lambda} \,
  \left( 1 - e^{- \lambda t} \right) \ . \label{loc}
\end{equation}
Thus, asymptotically the $A$ particles become localized, with $\langle 
{\vec x}(t)^2_A \rangle \to 2 d \, {\widetilde D} \, \rho_B(0) / \lambda$ in 
the limit $t \to \infty$.
On the other hand, if the $B$ density decays algebraically, see 
Eq.~(\ref{bds}), then for $0 < \alpha_B < 1$ our mean-field solution predicts
subdiffusive propagation (\ref{amd}) with $\alpha_A = \alpha_B$, whereas
asymptotically
\begin{equation}
  \langle {\vec x}(t)^2_A \rangle \sim {\widetilde D} \, \ln D t \label{2mf}
\end{equation}
if $\alpha_B = 1$.
Finally, for pair annihilation processes at the critical dimension $d_c = 2$,
governed by Eq.~(\ref{pa2}), one finds
\begin{equation}
  \langle {\vec x}(t)^2_A \rangle \sim {\widetilde D} \, (\ln D t)^2 \ ,
\label{2d2}
\end{equation}
and similarly we obtain for triplet annihilation in one dimension, see
Eq.~(\ref{ta1}),
\begin{equation}
  \langle x(t)^2_A \rangle \sim {\widetilde D} \, (D t \, \ln D t)^{1/2} \ .
\label{3d1}
\end{equation}

\section{Monte Carlo simulation methods}
\label{simul}

Our goal was to employ Monte Carlo simulations to compare $\alpha_A$ with
$\alpha_B$, as well as to determine the displacement probability distribution 
$n_A({\vec x},t)$ for the $A$ particles, and look for deviations from the 
Gaussian distribution (\ref{sol}) predicted by the mean-field approach.
The simulations discussed here were executed on a cubic lattice with periodic
boundary conditions in each spatial direction.
In one dimension the lattice contained between $10^4$ and $10^5$ sites, and
the two-dimensional lattice ranged in size from $100 \times 100$ to
$800 \times 800$.
In each simulation the system was initialized by putting one $B$ particle at
each site (initial density $\rho_B(0) = 1$ \cite{fnote2}), and by randomly
placing throughout the lattice a fixed number of $A$ particles with no site
exclusion.
For all of the data given below (except where otherwise noted), the density
of $A$ particles in the lattice was fixed at $\rho_A = 0.5$.
Each time step involved a complete update of the $A$ species, followed by a
complete update of the $B$ particles.
The simulation was terminated either when the number of $B$ particles reached
a certain lower limit (usually $0.1 \%$ of the number of lattice sites), or
after a fixed number of time steps (typically $\approx 10^6$).

With the exception of some particular runs in which the $B$ particles were
non-interacting (i.e., subject only to the decay $B \to \emptyset$) and could
thus be updated serially, we proceeded as follows.
Given $N$ particles at the beginning of the time step, $N$ random $B$ particles
on the lattice were chosen to be updated.
In a given time step, some $B$ particles might then be addressed more than once
while others not at all.
This is appropriate even though the number of $B$ particles is changing in 
time, because the net loss of $B$ particles per time step becomes less than one
on time scales short compared to the simulation length.
The $B$ particles were in general subjected to the reactions listed in
(\ref{rxs}), and a single update proceeded in that sequence.
To implement the process $B \to \emptyset$, the $B$ particle was deleted with
probability $\lambda$.
Next, the $B$ particle underwent an offspring reaction $B \to (m+1) \, B$ with
probability $\sigma$.
In the simulations discussed here, we chose $m = 1$, $2$, or $4$.
The offspring particle(s) were placed on the parent particle's nearest
neighboring sites such that no offspring were placed on identical places, and
were then subject to the annihilation reaction $n \, B \to \emptyset$ (with
$n = 2$, $3$, or $4$) if applicable.
If the $B$ particle being updated did not undergo an offspring reaction, it
subsequently hopped to a nearest neighbor site with some probability and was
then subject to annihilation, which required all $n$ $B$ particles to be
located on the same site.

Initially, the $A$ particles were also updated via Monte Carlo.
To this end, a random direction was chosen, and the $A$ particle was moved one
step in that direction provided at least one $B$ particle was present on the
destination site.
However, since the $A$ particles were independent, it was later determined
that they could be processed serially (i.e., by simply passing through the list
of $A$ particles).
Technically this represents a microscopically different method, since the Monte
Carlo procedure causes some particles to be updated more than once, and others
not at all at a given time step.
Yet we found that both variants produced identical macroscopic results.

Our main interest was to determine the asymptotic scaling behavior of the
global $B$ particle density in the lattice, as well as to measure the 
mean-square displacement of the $A$ species, both as functions of time $t$.
Henceforth, lengths will be measured in units of the lattice constant $a_0$, 
and time in units of Monte Carlo steps.
In most of our simulations we expect spatial inhomogeneity in the $B$ particle
distribution.
In particular, anti-correlations should develop in low dimensions for pure $B$
annihilation kinetics.
At the critical point for systems exhibiting phase transitions, at 
sufficicently large times the $B$ species distribution should become a
scale-free spatial fractal at length scales large compared to the lattice
constant and smaller than the system size (compare Fig.~\ref{fig:clus}).
Therefore we also periodically recorded the coordinates of both $A$ and $B$
particles in order to compute probability distributions and correlation
functions.
For example, the $B$ density correlation function is defined as 
\begin{equation}
  C_B({\vec x},t;{\vec x}',t') = \langle \rho_B({\vec x},t) \,
  \rho_B({\vec x}',t') \rangle - \langle \rho_B \rangle^2 \ . \label{cfn}
\end{equation}
By the translational and rotational invariance of the lattice, the equal-time
density correlations are really a function of $|{\vec x} - {\vec x}'|$ only.
At criticality, we have
\begin{equation}
  C({\vec x},t;{\vec x}',t) = C(|{\vec x} - {\vec x}'|) \sim 
  |{\vec x} - {\vec x}'|^{-2 \beta / \nu} \ , \label{ccf}
\end{equation}
whence we find at equal positions
\begin{equation}
  C({\vec x},t;{\vec x},t') = C(|t - t'|) \sim |t - t'|^{-2 \beta / z \nu} \ ,
\label{ctf}
\end{equation}
as a consequence of dynamic scaling.

To compute the equal-time correlation function (\ref{ccf}) in one dimension 
numerically, we fix a particular $B$ site and observe the distribution of all 
other $B$ particles as a function of the distance from it.
We measure this distribution for each $B$ particle fixed and then average the
resulting distributions to obtain $\langle \rho_B(x,t)\, \rho_B(x',t) \rangle$.
In higher dimensions, we use the lattice directions as representative of the
full distribution and compute $C(|{\vec x}-{\vec x}'|)$ for pairs 
$({\vec x},{\vec x}')$ with a common lattice coordinate in one direction.

In many situations we expected the measured quantities to be power laws as a
function of time $t$.
The simplest approach to computing the exponent $\alpha$ in a power law
relationship $\rho \propto t^{- \alpha}$ naturally is linear regression on
$\ln \rho$ vs. $\ln t$.
At the continuous phase transition separating active and absorbing states, one
expects such a power law dependence.
However, when the system is slightly above or below the critical parameters, it
usually behaves critically for some time before crossing over to super- or
subcritical behavior (either an exponential approach to the final $B$ particle
density, or a power law with a different exponent $\alpha'$).
If the system is not precisely at criticality (at least for the time scales
simulated), then offcritical behavior could lead to incorrect determination of
the critical exponent via linear regression.
However, one can also compute a local exponent $\alpha_b$ for a measured
quantity $\rho$ given by the expression \cite{perlsman02}
\begin{equation}
  \alpha_b = - \log_{b^2} \left[ \rho(b t) / \rho(t / b) \right]\ . \label{lex}
\end{equation}
Thus at time $t$, supposing a power law dependence on $t$, $\alpha_b$ is the
exponent inferred from the values of $\rho$ at $b \, t$ and $t / b$; i.e., 
these two data points define a line on a $\log$-$\log$ plot whose slope is
$- \alpha_b$.

The most time-consuming procedure in this study involved finding the critical
parameters for which the system was at criticality.
Typically, the parameters $\lambda$ and $\mu$ were fixed, and $\sigma$ was
varied.
The critical value of $\sigma$ was deduced by simultaneously increasing the
lower bound by definitely identifying systems as subcritical, and decreasing
the upper bound by characterizing supercritical systems.
In either case, the system behaved critically for some time (longer for 
$\sigma$ closer to the true critical value $\sigma_c$) before crossing over to 
its asymptotic behavior, and so critical power laws could be approximated from
the system's intermediate scaling behavior.
As noted previously, the $B$ species phase transition between active and
absorbing states induces a localization transition for the $A$ particles, with
critical subdiffusive behavior.
The phase transitions for both $A$ and $B$ particles clearly occur at the same
value of $\sigma_c$, and hence $\sigma_c$ can be determined independently by
measuring both the $B$ particle density decay and the $A$ species mean-square
displacements as a function of time.
We estimate our typical errors in determining critical exponents and the
subdiffusive $A$ species power laws to be $\approx \pm 0.01$.

We also attempted to improve the precision of our estimation of $\sigma_c$ by
a method suggested in Ref.~\cite{perlsman02}, which we now briefly describe.
For the measured quantity $\rho$, which at the critical parameter value
$\sigma_c$ depends only on some power of time $t$, i.e.,
$\rho(\sigma_c,t) \sim t^{- \alpha}$ with some exponent $\alpha$, one expects
in the critical regime $\sigma \approx \sigma_c$:
\begin{equation}
  \rho(\sigma,t) \simeq \rho(\sigma_c,t) \,
  \left[ 1 + c \, t^{1 / \nu_t} (\sigma - \sigma_c) \right] \ . \label{lap}
\end{equation}
Here $\nu_t = z \, \nu$ describes the critical slowing down as the control
paramater $\sigma$ approaches the phase transition at $\sigma_c$,
$\tau_c \sim |\sigma - \sigma_c|^{- \nu_t}$.
One first estimates $\sigma_c$, $\nu_t$, and $c$ from relatively short
simulations at various values of $\sigma$.
A simulation reaching large values of $t$ is then performed at the estimated
$\sigma_c$.
One obtains an improved estimate for $\sigma_c$ by replacing $\sigma$ and
$\rho(\sigma,t)$ in Eq.~(\ref{lap}) with the long simulation measurements, and
then finding the value of $\sigma_c$ for which $\rho(\sigma_c,t)$ is a straight
line.
This process may then be repeated to obtain the desired or computationally
accessible accuracy.
Note that inaccuracies in the estimates of $c$ and $\nu_t$ result in
second-order inaccuracies for $\sigma_c$ and $\alpha$ \cite{perlsman02}.
However, this method only works when $\sigma$ is already sufficiently close to
$\sigma_c$ so that this first-order approximation is valid.
In addition, statistical fluctuations in the data must be smoothed as much as
possible through averaging over multiple runs for each $\sigma$ value so that
$c$ and $\nu_t$ can be somewhat accurately determined.
Unfortunately, we found our simulations were not extensive enough in most cases
to apply this method and obtain even more reliable estimates of $c$, $\nu_t$,
and $\sigma_c$.

\section{Annihilation kinetics and anomalous diffusion}
\label{nihil}

We begin with the results of our Monte Carlo simulations for pure $B$ species
annihilation reactions.
These serve to test the mean-field description of the ensuing $A$ particle
anomalous diffusion in dimensions below, at, and above the critical dimension
$d_c(n)= 2/(n-1)$, and moreover provide a means to estimate the magnitude of
errors to be expected in our data.
In addition, the results for the $B$ pair annihilation model should describe
the subcritical behavior for the reactions exhibiting active to absorbing
transitions in the PC universality class.

\subsection{Spontaneous decay $B \to \emptyset$}
\label{Brad}

We first verified that our simulations correctly reproduced the $n=1$ solution
(\ref{mrd}) to the mean-field equation (\ref{mfb}) giving exponential $B$
density decay.
This result should be valid in any dimension since all particles evolve
independently.
The mean-field description for the $A$ particles then predicts their 
localization according to Eq.~(\ref{loc}).
We ran simulations in both one (system size $10000$) and two dimensions
(system size $100 \times 100$) using a decay rate $\lambda = 0.01$, and indeed
found excellent agreement at large times with the prediction (\ref{loc}).
Initially, however, the $A$ particles moved slower than suggested by 
Eq.~(\ref{loc}), consistent though with our different microscopic realization
of the reaction-controlled diffusion model:
the rules adopted in the simulations only distinguish between sites that are
occupied or unoccupied by $B$ particles, and so multiple site occupation
effectively corresponds to lower local densities $\rho_B$ in the analytical
description.
Yet at low $B$ particle densities multiple site occupation becomes negligible,
leading to $A$ species localization precisely as described by Eq.~(\ref{loc}).
We also computed the actual $A$ particle displacement distribution as a
histogram of final net displacements.
Using the measured average final mean-square displacement as input to
Eq.~(\ref{sol}) rather than estimating the coefficient ${\widetilde D}$, we 
found good agreement with the predicted Gaussian distribution in both one and 
two dimensions, although in $d = 1$ the data perhaps indicate a slight excess 
of particles localized very near their initial location.

In the results summarized above, at each Monte Carlo step the updated $B$
particles hopped to a nearest neighboring site with probability $1$.
We also investigated the effect of varying this probability.
Initially, the mean-square $A$ displacement then grows faster in situations
with low $B$ particle diffusivity because the probability of multiple $B$ site
occupation is reduced, leaving a larger fraction of available sites for the $A$
species.
However, as the $B$ particles are depleted the connectivity of the lattice
decreases, and the movement of the $B$ species quickly becomes the dominant
mechanism for $A$ particle diffusion.
We found an overall monotonic increase of final $A$ particle mean-square
displacements as a function of the $B$ diffusivity.
In one dimension the (temporary) localization of an $A$ particle requires only
two sites unoccupied by $B$ particles, compared with four sites in two 
dimensions.
Therefore the connectivity decreases more sharply in $d = 1$ as a function of
the $B$ density.
Consequently diminishing the $B$ species random walk probability has a more
pronounced effect in one dimension than in $d=2$.

\subsection{Pair annihilation $2 \, B \to \emptyset$}
\label{Bpan}

\begin{figure}
\includegraphics*[scale=0.55,angle=0]{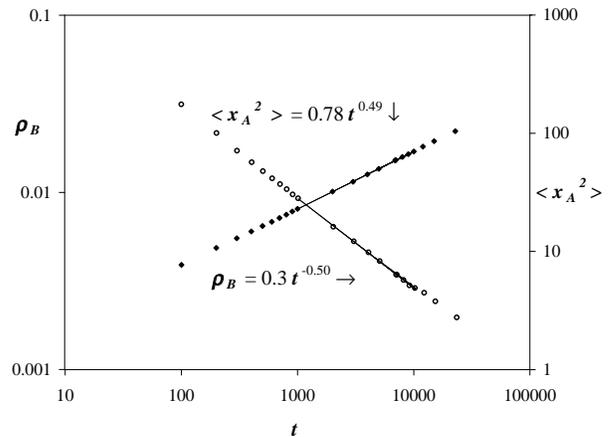}
\caption{\label{fig:pan1} The $B$ particle density and corresponding $A$ 
  species mean-square displacement for the pair annihilation reaction
  $2 \, B \to \emptyset$ with $\mu = 0.5$ in $d = 1$ demonstrating excellent 
  agreement with the predicted asymptotic power laws.}
\end{figure}

Since the critical dimension for the $B$ particle pair annihilation reaction is
$d_c(2) = 2$, the mean-field prediction (\ref{mas}) does not apply to either
the one- or two-dimensional simulations executed.
In $d = 1$ we expect according to Eq.~(\ref{pad}) $\rho_B(t) \sim t^{-1/2}$,
and using Eqs.~(\ref{m2s}) and (\ref{dpt}) the mean-field approach suggests 
$\langle x(t)_A^2 \rangle \sim t^{1/2}$.
Figure~\ref{fig:pan1} shows the simulation results (averaged over $20$ runs)
for $B$ pair annihilation in one dimension (system size $10000$) with
annihilation probability $\mu = 0.5$.
At first the $B$ density decays faster than Eq.~(\ref{pad}) predicts:
initially the effective power law should be close to the mean-field result
$\rho_B(t) \sim t^{-1}$ since the anti-correlations described in
Sec.~\ref{theor} develop only after some time has elapsed, whereupon the
asymptotic decay $\sim t^{-1/2}$ is approached.
Indeed, the graph verifies that the theoretical exponent $\alpha_B = 1/2$ is
adequately reproduced in the late time interval $10^3 < t < 10^4$.
We measured the corresponding exponent $\alpha_A$ for the mean-square $A$
particle displacement in this regime as well, and found that the simulation
results agree nicely with the mean-field exponent $\alpha_A = 1/2$, to the
precision obtained in our simulations.

We may also infer ${\widetilde D} \approx 0.69$ by matching our numerical
integral of $\rho_B(t)$ with $\langle x(t)^2_A \rangle$, though incomplete
knowledge of $\rho_B(t)$, particularly in the transient regime, introduces some
errors to this estimate.
A value ${\widetilde D} > 0.5$ indicates that more than one $A$ particle is
hopping to the same $B$ particle site on average.
When some $A$-$B$ correlations develop there may be a higher density of $A$
particles in the vicinity of a given $B$, which would in turn enhance the
average $A$ diffusion rate.

\begin{figure}
\includegraphics*[scale=0.6,angle=0]{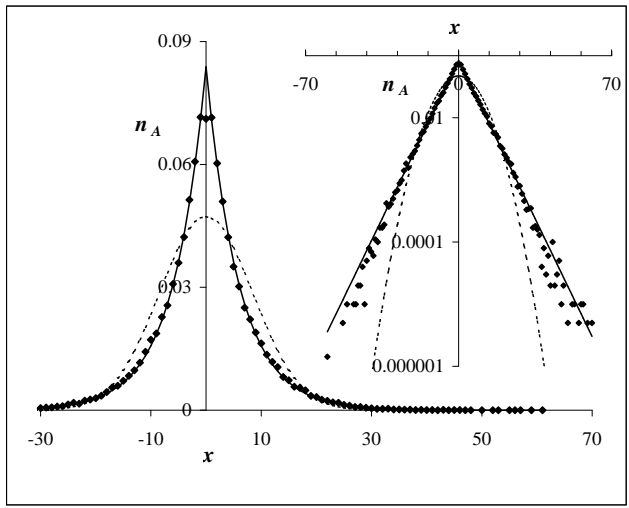}
\caption{\label{fig:pad1} The measured distribution of the $A$ particle
  displacements corresponding to the system shown in Fig.~\ref{fig:pan1} ($B$ 
  pair annihilation in one dimension, $\mu = 0.5$, measured at $t = 10000$).
  Both Gaussian (dashed) and exponential fits (full line), uniquely determined
  by normalization and fixing the second moment with the simulation data, are
  depicted, on linear and logarithmic (inset) scales.
  The exponential fit works very well, although the value at $0$ and perhaps
  the tails of the observed distribution appear smaller than in the fit.}
\end{figure}

Figure~\ref{fig:pad1} demonstrates that the predicted Gaussian distribution
Eq.~(\ref{sol}) is {\em not} observed, despite the agreement in the time 
dependence of the mean-square displacement resulting from the distribution.
Compared to a Gaussian with matching second moment, there is a distinct excess
of essentially localized particles (with very small displacements), which
necessarily implies longer `tails' at large displacement values.
A closer examination reveals that the simulation results for the $A$ particle
displacement distribution agree over a large range of displacements with the
normalized exponential distribution
\begin{equation}
  n_A(x,t) = \frac{1}{2 \, L(t)} \ e^{- |x| / L(t)} \ , \label{exd}
\end{equation}
where $L(t)$ denotes a time-dependent characteristic length scale.
As will be discussed below, the distribution (\ref{exd}) obeys dynamic scaling,
so we determined $L$ by matching the second moment of this function with the
simulation data at $t = 10000$.
As is evident from Fig.~\ref{fig:pad1}, the only significant deviations from
this fit appear for particles with zero displacement and at the tails of the
distribution, where simulation data contain greater error.

We can understand the qualitative features of this non-Gaussian distribution as
a result of the low connectivity of a one-dimensional lattice.
In small regions where all $B$ particles have been annihilated, the $A$ species
are temporarily localized and will have few subsequent chances to escape, thus
lowering their effective diffusivity.
Direct measurement of the $B$ pair correlation function, as well as the slow
$B$ density decay clearly indicate that particle anti-correlations have
developed.
These anti-correlations enhance the probability of finding substantial regions
of zero $B$ particle density.
However, as mentioned before in the discussion of the variation of the $B$
particle random walk probability in Sec.~\ref{Brad}, a significant part of $A$
movements probably result from hopping along with a particular $B$ particle
through several time steps (especially in $d = 1$).
Thus the effective diffusion coefficient is (at least temporarily) much higher
for such particles, and this effect contributes results in the longer `tails' 
in the displacement distribution.
We may think of a {\em local} diffusion rate $D({\vec x},t)$ which is
proportional to the distribution of available sites $b({\vec x},t)$.
Thus a particular particle will be subject to a temporally varying diffusion
rate that depends on its trajectory through the field $b({\vec x},t)$, recall
Eqs.~(\ref{lda}) and (\ref{noi}), just as the diffusing particle's trajectory
must be considered for understanding diffusion on a static fractal.
The distribution depicted in Fig.~\ref{fig:pad1} suggests a sort of phase
separation into different populations:
many particles are primarily subject to small diffusion rates, i.e., are mainly
localized in regions with low $B$ density, while some others experience quite
large diffusivities (by remaining in areas of high local $B$ particle 
densities). 

The variation of local $A$ diffusion rates is the result of inhomogeneities in
$b({\vec x},t)$ and the coupling that allows an $A$ particle to be carried
along by a particular wandering $B$ particle.
Yet our earlier mean-field description for the $A$ species displacement
distribution assumed that on average each $A$ particle evolves with an 
identical effective diffusion coefficient; i.e., the total number of hops for
each $A$ particle should be about the same.
However, spatial inhomogeneities over scales larger than the distances traveled
by typical $A$ particles will cause the effective diffusivities for a
particular time interval to vary spatially in a noticeable amount.
This effect should be especially pronounced in the case of the $B$ pair
annihilation process because the emerging anti-correlations will leave many
regions of space rather devoid of $B$ particles.
$A$ particles with larger effective diffusion coefficients probably follow a
particular $B$ particle for some time, since the $B$ anti-correlations render
them unlikely to be present in a region of high local $B$ density.
This distribution of $B$ density spatial fluctuations should persist to a large
extent throughout the simulation runs, which in turn should yield large
variations in the average diffusion coefficient experienced by the $A$ species.

We may construct a simple model to accommodate such variations as follows.
Suppose the number of $B$ particles, i.e., the number of available hopping
sites for the $A$ species, is decreasing proportional to some negative power of
time, as is on average often the case in our simulations.
If Eq.~(\ref{bds}) holds, we find for the fraction $f$ of $B$ particles that
disappear during a small time interval $\Delta t \ll t$:
$f(t) = \alpha_B \, \Delta t / t$.
Now assume that initially all of the $A$ particles are diffusing ordinarily
with an effective diffusivity $D$, and then after each small time interval
$\Delta t$, a fraction $f$ of the active $A$ particles becomes localized and
remains immobile for the remainder of the simulation.
This suggests that the $A$ particle distribution will be a superposition of
normalized Gaussians of increasing width but multiplied by a factor to be
found recursively from the condition that the $A$ particle number be conserved.
More precisely, in $d=1$ this distribution becomes (with initial time $t_0$ and
final time $t = t_M$, $M \geq 1$):
\begin{equation}
  P(x,t) = \sum_{k = 0}^{M} p(t_k) \, \frac{1}{(4 \pi D \, t_k)^{1/2}} \,
  \exp\left( - \frac{x^2}{4 \, D \, t_k} \right) \label{col}
\end{equation}
The prefactors $p(t_k)$ are then determined by keeping the integral of $P(x,t)$
normalized to unity: $\sum_{k=0}^M p(t_k) = 1$.
Recursively, one thus arrives at $p(t_0) = f(t_0)$,
$p(t_k) = f(t_k) \prod_{j=0}^{k-1} [1 - f(t_j)]$ for $1 \leq k \leq M-1$, and
at last $p(t_M) = \prod_{j=0}^{M-1} [1 - f(t_j)]$.
This picture assumes that the entire region surrounding an $A$ particle is
depleted of $B$ particles nearly simultaneously, and that this region is not
subsequently visited by other $B$ particles.
While this may be a decent approximation under the condition of $B$ particle
anti-correlations, it is rather more difficult to justify in cases of emerging
positive correlations (as we shall discuss below when offspring reactions are
introduced).
However, even then some $B$ clusters are eventually eliminated, leaving the
nearby $A$ particles localized, at least temporarily.
Furthermore the active $A$ particles are diffusing normally with
$\langle x(t)^2_A \rangle \sim t$ in their local $B$ cluster until they become
trapped.

\begin{figure}
\includegraphics*[scale=0.55,angle=0]{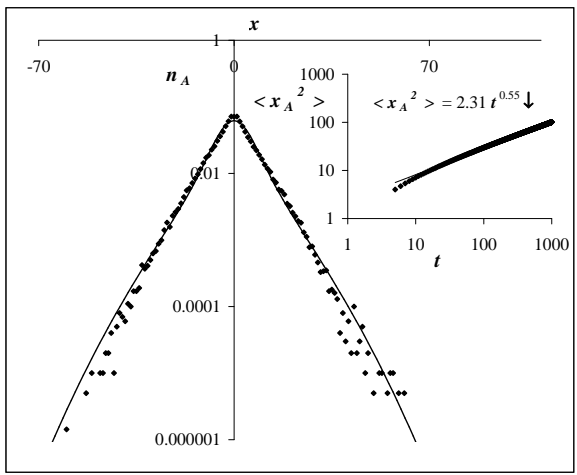}
\caption{\label{fig:clm1} The results from the `continuous localization' model
  superimposed with the measured distribution of the $A$ particle displacements
  corresponding to the system shown in Fig.~\ref{fig:pan1}.
  The model reproduces the faster than exponential drop-off in the tails of the
  distribution, but still underestimates the fraction of highly localized
  particles.
  The inset depicts the power law increase of $\langle x(t)^2_A \rangle$ as
  computed from the model.}
\end{figure}

Applying this simplified model to the pair annihilation system, we set
$\alpha_B = 0.5$, and we have chosen $D = 0.2$.
We then summed Gaussian distributions for simulation times ranging from $t=5$
to $100$, localizing a fraction $f(t) = \alpha_B \, \Delta t / t$ at each
integer $t$.
The result is depicted in Fig.~\ref{fig:clm1}, after appropriately rescaling
the axes to match normalization and the measured second moment of the
simulation data.
Apart from large deviations for small displacements, this `continuous
localization' fit appears to agree remarkably well with the simulation results.
Indeed this generated $A$ displacement distribution even begins to decrease
more quickly near its tails, as may also be observed in the simulation data.
The disagreement between the data and model may of course be traced to the
simplifications assumed in its construction.
However, the error also might arise from choosing the initial distribution:
at the beginning of the simulation the number of $B$ particles decreases
sharply before anti-correlations have developed, and the above estimate for the
fraction of localized $A$ particles is not valid for $\Delta t/t \sim 1$.
Despite its shortcomings, the model seems to capture most of the features we
have observed in the $A$ species displacement distribution.
We also computed the time dependence of the mean-square displacement using the
`continuous localization' model.
Following a transient behavior, the inset in Fig.~\ref{fig:clm1} shows
$\langle x(t)^2_A \rangle \sim t^{0.55}$, close to the expected power law with
$1 - \alpha_A = 0.5$ (which we also observed in the simulation).
By trying a variety of $\alpha_B$ values one can generate different
approximate power laws, but the measured subdiffusive $A$ displacement
exponent was always found to be slightly larger than $1 - \alpha_B$.

\begin{figure}
\includegraphics*[scale=0.55,angle=0]{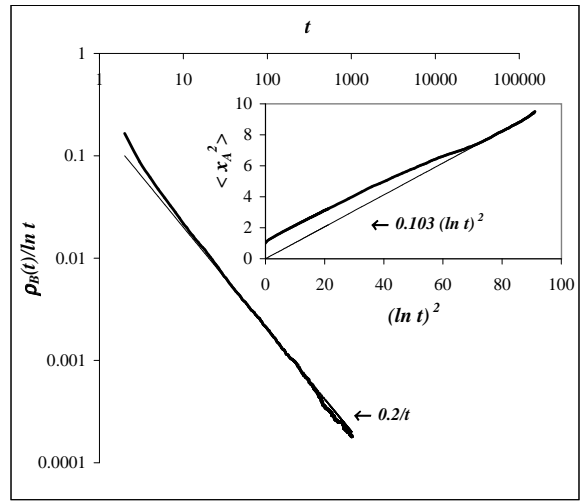}
\caption{\label{fig:pan2} The $B$ particle density and corresponding $A$
  species mean-square displacement for the pair annihilation reaction
  $2 \, B \to \emptyset$ with reaction probability $\mu = 1$ in $d = 2$.
  At long times, both plots indicate the expected logarithmic corrections to
  the mean-field scaling laws.}
\end{figure}

To examine the non-Gaussian character of $n_A(x,t)$ further, we measured its 
higher moments as a function of time.
In $d = 1$, Eq.~(\ref{mgm}) yields for the even moments of the mean-field
Gaussian distribution:
$\langle x(t)^{2k}_A \rangle = (2k - 1)!! \, \langle x(t)^2_A \rangle^k$.
While this factorization of higher moments still holds to the accuracy of our
simulation data, we found the prefactors
$c_k = \langle x(t)^{2k}_A \rangle / \langle x(t)^2_A \rangle^k$ in our
measurements to differ from the predicted values $c_2 = 3$, $c_3 = 15$, and
$c_4 = 105$:
we measured the considerably larger values $c_2 \approx 5.5$, $c_3 \approx 55$,
and $c_4 \approx 1100$.
These numbers are actually closer to the values one would obtain from the
exponential distribution (\ref{exd}), namely $c_k = (2k)! / 2^k$, i.e.,
$c_2 = 6$, $c_3 = 90$, and $c_4 = 2520$, but still off by a factor of about two
for the higher moments.
Thus the exponential fit cannot entirely describe the measured distribution
either.
Nevertheless, the factorization property
$\langle x(t)^{2k}_A \rangle \propto \langle x(t)^2_A \rangle^k$ found in the
simulation data is significant, for it indicates {\em dynamic scaling}:
once the time dependence of the characteristic length scale
$L(t) = \langle x(t)_A^2 \rangle^{1/2} \sim t^{(1 - \alpha_A)/2}$ is factored
out, the shape of the probability distribution should remain constant in time.
The distributions (\ref{sol}) and (\ref{exd}) clearly display this feature.
Since the probability distribution is fully determined by all its moments, our
measurements of the first four moments indicate that the true $n_A(x,t)$ obeys
dynamic scaling as well.

\begin{figure}
\includegraphics*[scale=0.55,angle=0]{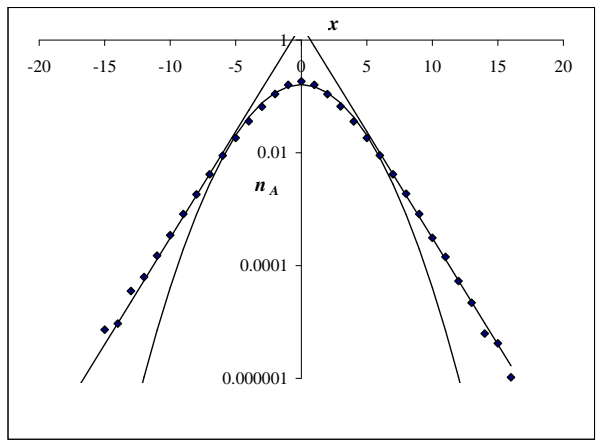}
\caption{\label{fig:pad2} The measured distribution of the final $A$ particle
  displacements corresponding to the system shown in Fig.~\ref{fig:pan2}
  ($B$ pair annihilation in two dimensions, $\mu = 1$), shown with Gaussian and
  unnormalized exponential fits.}
\end{figure}

At $d_c(2) = 2$ one obtains the typical logarithmic corrections to the
mean-field result, viz. Eq.~(\ref{pa2}) for the $B$ particle density, and
Eq.~(\ref{2d2}) for the $A$ species mean-square displacement.
Simulations were also carried out at various values of $\mu$ for a
two-dimensional system.
Figure~\ref{fig:pan2} illustrates the apparent agreement with the above
predictions at long times for simulations run with annihilation probability
$\mu = 1$ (averaged over $5$ runs on a $100 \times 100$ lattice).
However, such a large annihilation rate severely suppresses the $A$ particle
diffusion, so one should not really draw too firm conclusions from these
measurements.
We have again measured the corresponding distribution of the $A$ particle
displacements.
In Fig.~\ref{fig:pad2} we see that $n_A({\vec x},t)$ is non-Gaussian and
resembles its one-dimensional counterpart (Fig.~\ref{fig:pad1}), though the
deviations from Eq.~(\ref{sol}) are less pronounced.
When $\mu \ll 1$, anti-correlations develop over much larger time scales and as
the corrections are only logarithmic, for $\mu = 0.01$ ($20$ runs on a
$200 \times 200$ lattice) we merely recovered the mean-field results in the
regime accessible to our simulations.
Though not shown here, the $B$ density decay was captured by the mean-field
power law on the time scales of our runs, and the $A$ particle mean-square
displacement followed the integral of that quantity after some initial
transient behavior.
Accordingly, the $A$ particle distribution was essentially Gaussian in this
situation.

\subsection{Triplet annihilation $3 \, B \to \emptyset$}
\label{Btan}

\begin{figure}
\includegraphics*[scale=0.53,angle=0]{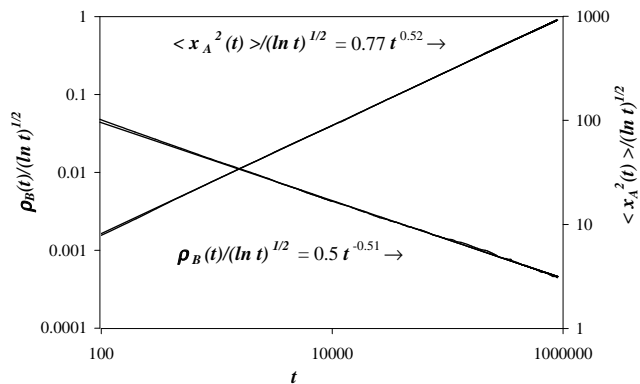}
\caption{\label{fig:tan1} The $B$ particle density and corresponding $A$
  species mean-square displacement for the triplet annihilation reaction
  $3 \, B \to \emptyset$ with $\mu = 1$ in $d = 1$, both displaying the
  expected logarithmic corrections to the mean-field scaling laws.}
\end{figure}

Next we consider the triplet annihilation reaction $3 \, B \to \emptyset$.
The upper critical dimension here is $d_c(3) = 1$.
Hence in one dimension we expect logarithmic corrections to the mean-field
power law $B$ density decay, as given in Eq.~(\ref{ta1}).
According to the simple mean-field picture, this leads to Eq.~(\ref{3d1})
for the $A$ species mean-square displacement.
Figure~\ref{fig:tan1} shows our simulation results (averaged over $20$ runs
on a lattice with $10000$ sites, annihilation probability $\mu = 1$).
We are able to clearly detect the logarithmic corrections even at $\mu = 1$
since this reaction is a much slower process than pair annihilation (requiring 
three particles to meet on a site).
The power law regression on $\langle x(t)_A^2 \rangle / (\ln t)^{1/2}$
yields a value of $1 - \alpha_A \approx 0.52$, whereas the expected value is
$0.50$.
Supposing the mean-field result fully applies here, this gives an idea of
the overall precision of our simulation data.
The $A$ particle displacement distribution also agrees well with the predicted
Gaussian, apart from a slight excess of presumably localized particles around
$\langle x_A^2 \rangle = 0$ and correspondingly longer `tails' of the
distribution.
Yet the deviation is much smaller than in the pair annihilation case because
the anti-correlations are less pronounced here.

\begin{figure}[b]
\includegraphics*[scale=0.6,angle=0]{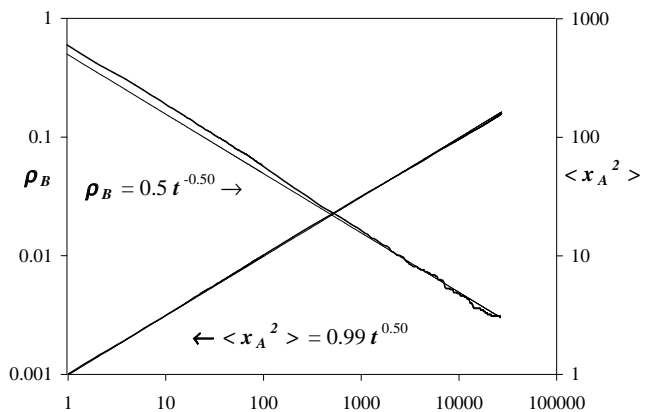}
\caption{\label{fig:tan2} The $B$ particle density and $A$ species mean-square 
  displacement for the triplet annihilation reaction $3 \, B \to \emptyset$ 
  ($\mu = 1$) in $d = 2$, confirming the mean-field power laws.}
\end{figure}

For $d = 2 > d_c$, the mean-field result (\ref{mas}) should provide a correct
description, i.e., for $n = 3$ in the long-time limit $t \gg \tau_3$:
$\rho_B(t) \sim t^{-1/2}$ and $\langle {\vec x}(t)_A^2 \rangle \sim t^{1/2}$.
Figure~\ref{fig:tan2} demonstrates agreement with these mean-field predictions
after $10^3$ or fewer timesteps (from $3$ runs on a $100 \times 100$ lattice
with again $\mu = 1$).
In fact, to our resolution the mean-square displacement of the $A$ species
converges to the mean-field power law markedly faster than the $B$ particle
density.
As yet we have no explanation for this surprising observation.

\subsection{Quartic annihilation $4 \, B \to \emptyset$}
\label{Bqan}

The kinetics of quartic annihilation should be aptly described by the
mean-field rate equation (\ref{mfb}) in all physical dimensions since
$d_c(4) = 2/3$.
Setting $n = 4$ in Eq.~(\ref{mas}) we thus expect $\rho_B(t) \sim t^{-1/3}$ for
sufficiently large $t \gg \tau_4$.
According to Eqs.~(\ref{m2s}) and (\ref{dpt}),
$\langle \vec x(t)_A^2 \rangle \sim t^{2/3}$.
We ran simulations in two dimensions on a $100 \times 100$ lattice (with
$\mu = 1$, averaged over $8$ runs) and found good agreement with these
mean-field scaling laws at sufficiently long times both for the $B$ particle
density decay and the $A$ species mean-square displacement.
For the annihilation processes, larger $n$ values imply longer crossover
time scales $\tau_n$ since $n$ particles must meet for a reaction to occur, see
Eq.~(\ref{mas}).
Indeed, by comparing with the results for the triplet reactions, we noticed
that the time interval of transient behavior prior to convergence to the
asymptotic mean-field power law was at least an order of magnitude longer for
the $n = 4$ system.
As in the triplet simulations, the convergence to mean-field behavior occured
faster for the $A$ particle mean-square displacement than for the total $B$
density.

\section{Active to absorbing state phase transition and localization}
\label{trans}

After this preliminary study with pure $B$ annihilation kinetics, we now turn
our attention to the primary goal of our investigation, namely anomalous
diffusion on a dynamic fractal (see Fig.~\ref{fig:clus}).
Each reaction discussed below exhibits a continuous phase transition separating
active and absorbing stationary states.
At the critical point, the $B$ particle distribution is known to be fractal in
space and time.
Thus the assumption of a homogeneous $B$ particle distribution is significantly
violated, and we expect the mean-field description for the $A$ species
propagation to be inadequate.
Our aim has been to measure and characterize the deviations from the mean-field
predictions.
All of the phase transitions in the $B$ particle system to be discussed here
are described either by the directed percolation (DP) or parity conserving (PC)
universality classes.
The accepted critical DP exponents are listed in Tab.~\ref{exptb}.
As mentioned in the Sec.~\ref{intro}, a PC phase transition is observed in
$d = 1$ when $\lambda = 0$ in the reaction scheme (\ref{rxs}), and $m$ is even;
in higher dimensions the (degenerate) critical point occurs at zero branching
rate and is governed by mean-field exponents \cite{cardy96}.
The PC exponents in $d = 1$ are also given in Tab.~\ref{exptb}.

Knowledge of the $B$ particle density behavior of these systems in the active
and absorbing states near the phase transition is also important, both in
locating the critical point and for comparison with the asymptotic critical
scaling laws.
In fact, in simulations we will inevitably always be slightly away from the
precise critical control parameter values.
However, in the vicinity of the phase transition we expect the system to
exhibit the critical power laws for some time interval before crossing over to
sub- or supercritical behavior.
The closer the system parameter values are to the critical ones, the longer
this critical regime lasts, as also suggested by Eq.~(\ref{lap}), and the more
precisely critical exponents can be measured.
When $m$ is odd (independent of $\lambda$), we expect systems in the absorbing
state (i.e., dominated by annihilation reactions) to exhibit behavior dictated
by Eq.~(\ref{mrd}), the solution to $B$ particle radioactive decay, with some
effective rate $\lambda_{\rm eff}$, dependent on $\lambda$, $\sigma$, and 
$\mu$.
To be more precise, let us consider the reactions (\ref{rxs}) with $m = 1$ and
$n = 2$ as is the case in many of the situations examined below.
Recalling that the branching reactions are local (offspring particles are
placed on the parents' neighboring sites), in low dimensions there is a
significant probability that the parent and offspring will meet in the next few
time steps after the branching reaction and undergo pair annihilation.
Together, the branching and subsequent annihilation reactions generate the
decay $B \to \emptyset$, even if $\lambda = 0$ on the outset.
Generally, this is true when $n = 2$ and $m$ is odd.
However, when $n = 2$ and $m$ is even, local parity conservation eliminates the
possibility to generate `spontaneous' death processes.
Thus in the PC universality class the inactive phase is characterized by the
power laws of the pair annihilation reactions $2 \, B \to \emptyset$, whence in
subcritical systems one should asymptotically observe the behavior given in
Eq.~(\ref{pad}) with $d = 1$.
In the active phases of both DP and PC systems, the $B$ density will approach
its stationary value exponentially.
This follows immediately from linearizing the corresponding mean-field rate
equations.

\subsection{Directed percolation universality class, $d = 1$}
\label{DP1}

\begin{figure}
\includegraphics*[scale=0.55,angle=0]{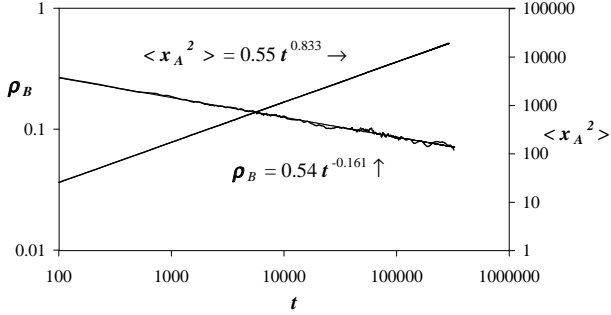}
\caption{\label{fig:dpc1} The $B$ particle density and $A$ species mean-square
  displacement for the DP reactions $B \to \emptyset$ ($\lambda = 0.01$),
  $B \to 2 \, B$ ($\sigma = 0.8975$), $2 \, B \to \emptyset$ ($\mu = 1$) in 
  $d = 1$.}
\end{figure}

\begin{figure}[b]
\includegraphics*[scale=0.5,angle=0]{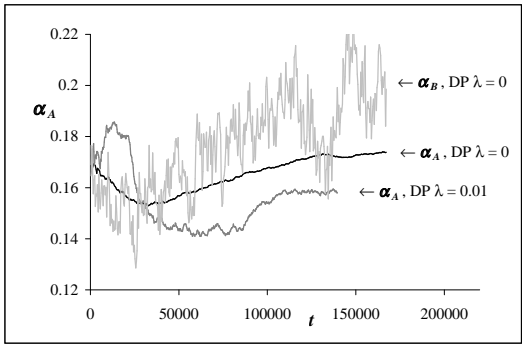}
\caption{\label{fig:dpc1a} The local exponent $\alpha_A$ corresponding to data
 shown in Fig.~\ref{fig:dpc1} (with $b = 2$) and Fig.~\ref{fig:dpc0}
 (with $b = 4$).}
\end{figure}

To search for the DP transition in one dimension, we considered the reactions
(\ref{rxs}) with $m=1$ and $n=2$, with $\lambda = 0.01$ and $\mu = 1$ fixed
while $\sigma$ was varied.
Notice that setting $\mu = 1$ eliminates the possibility of multiple $B$
particles occupying the same site, so that the $B$ particle density exactly
corresponds to the density of available lattice sites for $A$ particle hopping.
According to the list in Tab.~\ref{exptb}, $\rho_B(t) \sim t^{-0.159}$ at
criticality, which implies within the mean-field approximation
($\alpha_A = \alpha_B$) that $\langle x(t)_A^2 \rangle \sim t^{0.841}$.
Our closest estimate of the critical point is $\sigma_c \approx 0.8975$.
Figure~\ref{fig:dpc1} shows the power law dependence of $\rho_B(t)$ and
$\langle x(t)_A^2 \rangle$, obtained from $26$ runs on a lattice with $10000$
sites.
The double-logarithmic linear regressions yield values $\alpha_B = 0.161$ and
$\alpha_A = 0.167$ over more than three decades, i.e.,
$\alpha_A - \alpha_B \approx 0.006$.
As we estimate the numerical uncertainty of the measured exponents to be at 
least $\pm 0.01$, we observe agreement both between $\alpha_B$ and the DP 
prediction as well as between $\alpha_B$ and $\alpha_A$ within our error bars.
We also computed local exponents, as defined in Eq.~(\ref{lex}) setting $b=2$.
Figure~\ref{fig:dpc1a} indicates that after following some transient behavior, 
$\alpha_A \approx 0.158$, in excellent agreement with the critical DP value.
Thus we conclude that our mean-field prediction of the $A$ particle mean-square
displacement time dependence works excellently for this reaction at least on
the time scales we were able to access with our simulations.

\begin{figure}
\includegraphics*[scale=0.55,angle=0]{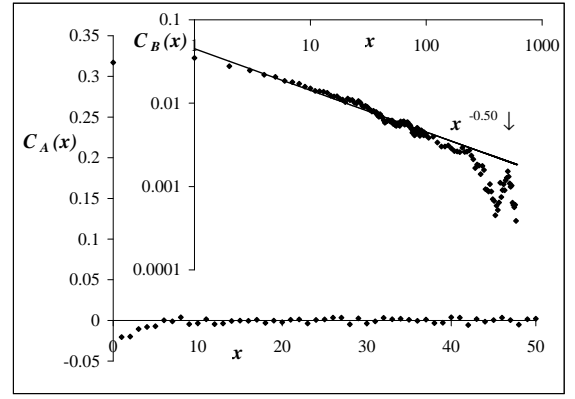}
\caption{\label{fig:dpf1} The $A$ and $B$ particle pair correlation functions
  $C_A(x)$ and $C_B(x)$ (inset) at the DP critical point ($d = 1$, measured at
  $t=50000$).
  $C_B(x) \sim |x|^{-2 \beta/\nu}$ is approximately scale-invariant.
  The $A$ particles are very likely to be found on the same site, but there is
  a slight negative correlation with nearby sites.
  $C_A(x)$ is essentially zero at distances $|x| > 10$.}
\end{figure}

In Fig.~\ref{fig:dpf1} (inset) we demonstrate the algebraic dependence of the
$B$ particle correlation function on particle separation, indicating the
spatially fractal structure at criticality.
The data were taken at $t = 50000$, averaged over $9$ simulations.
The measured exponent $2 \beta / \nu$ agrees well with the DP prediction $0.50$
in $d = 1$ over two decades.
We remark that for small distances we observed that the correlation function 
scales as $C_B(|x|) \sim |x|^{-\beta/\nu}$, i.e., with precisely one half the
asymptotic scaling exponent.
The origin of this can be understood as follows: at low $B$ particle densities,
the site occupation number $n(x)$ is either zero or one; hence for $|x| < \xi$
$\langle n(x) \, n(0) \rangle \approx \langle n^2 \rangle = \langle n \rangle
\sim \rho_B \sim \xi^{-\beta/\nu}$.
The proportionality factor here must be a scaling function $f(r/\xi)$, and
demanding that the $\xi$ dependence must cancel at criticality, we obtain the
aforementioned scaling law.
As we are not precisely at the critical point, we see a crossover to
exponential decay at large distances.

\begin{figure}
\includegraphics*[scale=0.55,angle=0]{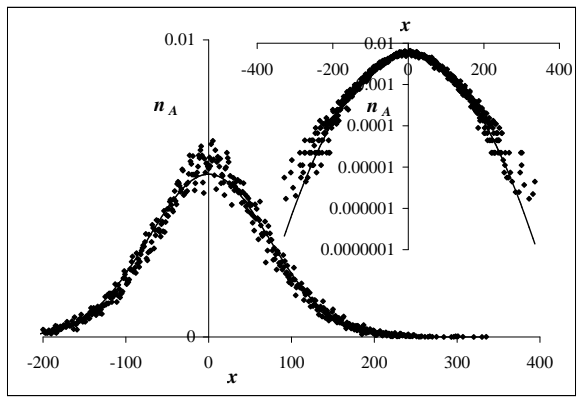}
\caption{\label{fig:dpd1} The measured distribution of the $A$ particle
  displacements corresponding to the system shown in Fig.~\ref{fig:dpc1} (DP 
  critical point for the $B$ system in $d = 1$, measured at $t = 50000$).}
\end{figure}

Figure~\ref{fig:dpd1} depicts the measured full $A$ particle displacement
distribution at $t = 50000$.
It displays an excess of localized particles and corresponding long tails of
the distribution, similar to the pair annihilation case.
But the deviation from a Gaussian distribution appears much less pronounced in
this system.
The power law dependence of $B$ particle spatial correlations qualitatively
suggests the observed distribution.
Such strong correlations yield macroscopic regions of relatively large $B$
densities which an $A$ particle may traverse and thus acquire a large
displacement.
However, such clusters necessarily indicate compensating large regions of very
low $B$ particle density, where the $A$ particles become highly localized.
The $A$ particles at the fractal `boundary', which by its nature accounts for
a significant fraction of the system volume, appear to behave in accord with
the mean-field prediction.
This is suggested by the fact that the data in Fig.~\ref{fig:dpd1} agree well
with a Gaussian distribution for intermediate displacement values.

Yet it was in these fractal boundaries that we expected the deviation from the
mean-field approximation to arise.
One possibility we had imagined is that $A$ particles in the boundary region
are driven towards denser $B$ regions owing to the particle density gradient
expected from the formation of clusters, thus destroying the spatial
homogeneity of the $A$ particles and resulting in a higher net diffusion rate
than assumed by mean-field theory.
This behavior can be distinguished by computing the $A$-$A$ particle
correlations.
If the described large-scale $A$-$B$ coupling were strong enough, then we would
expect to see some $A$ particle correlations as they congregate in regions of
high $B$ density.
But these correlations were not observed, as demonstrated in
Fig.~\ref{fig:dpf1}.
We may attribute the inability of the $B$ particles to induce significant
correlation in the $A$ system to the low connectivity of a one-dimensional
lattice.
However, $C_A(0)$ does indicate a significantly increased probability to find
multiple occupation of a given site.
A weak negative correlation seems to have developed for $0 < |x| < 5$.
Such a distribution may be the result of the `piggy-back' effect discussed
earlier:
several $A$ particles may in fact follow a single $B$ particle through a number
of time steps.
If the $B$ particle is subsequently annihilated, the `piggy-backing' $A$
particles are all (temporarily) localized at their current site.
The negative correlation may just be the compensation for the effective mutual
attraction of a group of nearby $A$ particles all following the same $B$.

We have also measured the higher moments of the $A$ displacement distribution
$n_A(x,t)$.
We found $\langle x(t)^{2k}_A \rangle = c_k \, \langle x(t)^2_A \rangle^k$,
indicating dynamic scaling, albeit with $c_2 = 3.3$, $c_3 = 20$, and
$c_4 = 185$ compared with the values $3$, $15$, and $105$ that would result
from a Gaussian.
A few distinctions between the pair annihilation case and this DP phase
transition may account for the significant differences in the corresponding
deviations from the mean-field $A$ displacement distribution.
Firstly, the $B$ particle density decays much slower in the DP case
($\alpha_B \approx 0.16$ as compared with $0.5$).
This slower decay may cause the DP distribution to be dominated by the `active'
$A$ particles.
For a fixed system size we will also observe the DP system for longer
durations.
Since the $B$ particles undergo ordinary diffusion while reacting, these
extended time scales imply that previously localized regions (where the local
$B$ density is zero) are likely to be visited by wandering $B$ particles.
Thus the localized portion of the $A$ particle distribution becomes smeared as
these $A$ particles are reactivated, which makes it more likely that all $A$
particles experience roughly the same average effective diffusion coefficient
throughout the simulation (as assumed in the mean-field approach).
Finally, at the DP phase transition, the $B$ particles are positively
correlated, whereas the pair annihilation process induces negative
correlations.
Positive correlations may also contrive to weaken particle localization, since
the $A$ species in regions of high local $B$ density are less dependent on
single $B$ particles for their mobility.
We also note that as a consequence of the effective `slaving' of the $A$
species by the $B$ particles, the correlation length exponents $\nu$ and
$\nu_t = z \nu$ should be identical for both the active to absorbing transition
in the $B$ system and the induced $A$ localization transition.
Indeed, to the accuracy we could determine those exponents by means of
Eq.~(\ref{lap}) and the method described in Ref.~\cite{perlsman02}, this
equality appeared to hold.

The expected time independence of the equal-time $B$ density correlation
function and the extent of the $B$ particle clusters explain the time
independence of the $A$ displacement distribution.
Recall that the origin of the non-Gaussian distribution is that the $A$
particles experience different effective diffusivities depending on their
location with respect to $B$ clusters.
What determines an $A$ particle's placement in the distribution is the average
diffusion coefficient experienced over the length of the simulation, neglecting
the possibility of biased diffusion due to $B$ particle gradients.
For the shape of the displacement distribution to remain constant in time, the
shape of the distribution of time-averaged effective diffusion rates most 
likely remains constant as well (even while on average the effective $A$
species diffusion rate is decreasing with time).
Subsequent measurement of the $A$ diffusivity distribution supports this
interpretation, see Fig.~\ref{fig:difr} in Sec.~\ref{concl} below.
This is only possible if $A$ particles are able to remain in a dense $B$ region
for a large portion of the simulation, made possible by large cluster sizes
implied by the power law correlation function, while others remain in a low
density region for long durations.
These two persisting extremes maintain the non-Gaussian nature of the $A$
species displacement distribution.
However, $A$ particles at the fractal boundary, which comprises a considerable
volume of the system and thus contains a large fraction of the $A$ particles,
may all see roughly the same effective diffusion rate over the length of the
simulation, which results in the Gaussian mid-region of $n_A(x,t)$.
Furthermore, the stationarity of the shape of the distribution of effective $A$
diffusivities is likely the reason we see such good agreement between
$\alpha_A$ and $\alpha_B$.
For instance, even if the global average instantaneous diffusion rate evolved
as ${\widetilde D} \langle \rho_B(t) \rangle$ (as we should expect), a changing
distribution shape, such as increasing enhancement of the phase separation
between `localized' and `active' regions while maintaining the proper global
diffusion rate, would induce additional time dependence and cause deviations
from Eq.~(\ref{m2s}).
However, we have no precise argument for the apparent time-independence of
these distributions.

\begin{figure}
\includegraphics*[scale=0.55,angle=0]{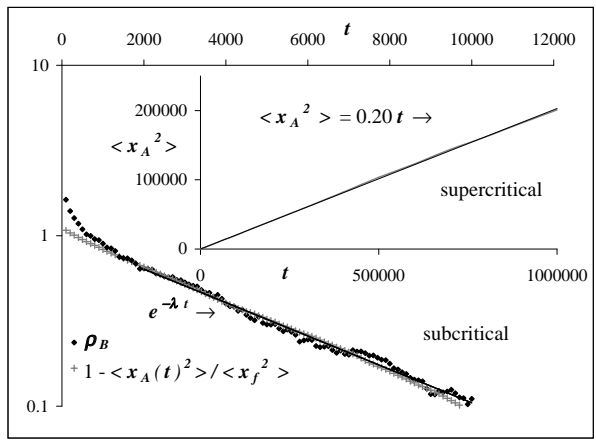}
\caption{\label{fig:dps1} $A$ species mean-square displacement in the super- 
  and subcritical regimes for the DP processes of Fig.~\ref{fig:dpc1} ($d=1$).
  For $\sigma = 0.910 > \sigma_c$, one finds normal $A$ diffusion.
  In the subcritical regime ($\sigma = 0.890$), the $B$ particle density decays
  exponentially with $\lambda_{\rm eff} \approx 0.0002$, and the mean-square
  $A$ displacement approaches the asymptotic value exponentially with the same
  time constant.
  (Both subcritical data sets are rescaled to a unit prefactor.)}
\end{figure}

We also sought to verify the predicted behavior away from the phase transition.
Setting $\sigma = 0.910$, sufficiently above the critical branching rate, we
observed almost immediate convergence of the $B$ particle density to its active
state saturation value $\rho_s = \langle \rho_B(t) \rangle \approx 0.26$, with
standard deviation $0.009$.
As expected and depicted in the inset of Fig.~\ref{fig:dps1}, the $A$ species 
then exhibits normal diffusion ($\alpha_A = 0$).
(The slight deviation at the end of the single run on 10000 sites may be a
finite-size effect.)
By Eqs.~(\ref{m2s}) and (\ref{dpt}) we also expect the prefactor to be
$2 {\widetilde D} \rho_s$, wherefrom we estimate ${\widetilde D} \approx 0.39$.
For ordinary diffusion, in the absence of any correlations,
${\widetilde D} = 0.5$.
Indeed, our algorithm dictates choosing a hopping direction first, and then
checking for the presence of a $B$ particle at the destination site.
Thus on average one of two neighboring $A$ particles will hop to a $B$ site.
To observe subcritical behavior, we set $\sigma = 0.890$.
Figure~\ref{fig:dps1} illustrates the convergence to the expected exponential
density decay with an effective decay rate $\lambda_{\rm eff} \approx 0.0002$
($10$ runs on a system with $10000$ sites).
Initially (for $t < 1000$), a brief quasi-critical regime is visible with
power-law decay $\rho_B(t) \sim t^{-0.26}$ and correspondingly
$\langle x(t)_A^2 \rangle \sim t^{0.77}$.
For $t > 2000$ we then observe exponential convergence to $\rho_B = 0$ and the
stationary value for $\langle x(t)_A^2 \rangle$, both with the same time
constant $\lambda_{\rm eff} \approx 0.0002$.

\begin{figure}
\includegraphics*[scale=0.55,angle=0]{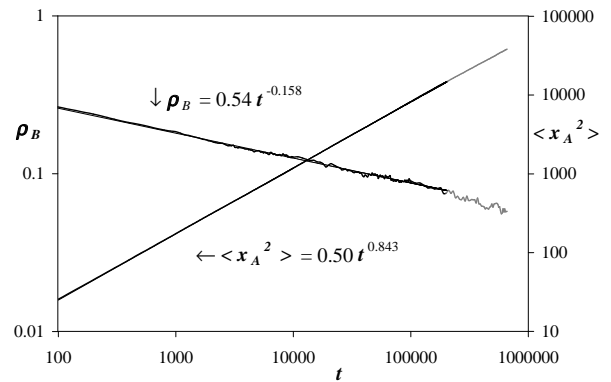}
\caption{\label{fig:dpc0} The $B$ particle density and $A$ species mean-square 
  displacement for the DP reactions (branching and annihilating random walks 
  with odd offspring number) $B \to 2 \, B$ ($\sigma = 0.8930$) and
  $2 \, B \to \emptyset$ ($\mu = 1$) in $d = 1$.
  For $t < 200000$, the critical power laws are observed.
  Subsequently a crossover to subcritical behavior can be seen.}
\end{figure}

As discussed above, we expect the macroscopic properties of the DP phase
transition to be independent of the value of $\lambda$, for $d \leq 2$ even
persisting to $\lambda = 0$, i.e., the absence of spontaneous decay.
To check this, we ran simulations for the reactions (\ref{rxs}) with $m=1$ and
$n=2$, setting $\lambda = 0$, $\mu = 1$, and varying $\sigma$.
We estimated $\sigma_c \approx 0.8930$, slightly below the value found when
$\lambda = 0.01$.
Figure~\ref{fig:dpc0} demonstrates the critical behavior deduced from data
taken over more than three decades (20 runs in a lattice with 20000 sites),
before the transition to subcritical behavior becomes evident.
We found $\alpha_B \approx 0.158$, in superb agreement with the expected DP
value $0.159$.
We measured $\alpha_A = 0.157$, so $\alpha_B - \alpha_A \approx 0.001$.
Again any significant deviations from the mean-field prediction at the critical
point, if present at all, must arise at time scales inaccessible to our
simulations.

However, the local exponents $\alpha_A$ and $\alpha_B$, setting $b = 4$ in
Eq.~(\ref{lex}), turned out not to be constant.
In the regime from which we inferred the global values of $\alpha_B$, the local
values oscillate about their averages by $\approx 0.01$, whereas the local
$\alpha_A$ is a much smoother function in time (see Fig.~\ref{fig:dpc1a}).
The steady increase in $\alpha_A$ (and $\alpha_B$) indicates that the system is
actually in the inactive phase.
We found the increase in the local $\alpha_B$ exponent to be significantly
faster.
However, this is not an indication of deviation from the mean-field prediction
for the mean-square displacement of the $A$ species.
We in fact computed numerically the integral of the $B$ particle density and
compared it with the measured mean-square $A$ displacement.
We found that for $t > 10000$, the error between the two measurements was less
than $1\%$.
We estimated ${\widetilde D} = 0.39$ for this reaction in order to obtain the
best match between the measured mean-square $A$ displacement and the integral
of $\rho_B(t)$.
This value is in good agreement with estimates from other reactions.
Both the $B$ and $A$ particle correlation functions and the $A$ species
displacement distribution were basically identical to Figs.~\ref{fig:dpf1} and
\ref{fig:dpd1}.

\subsection{Directed percolation universality class, $d = 2$}
\label{DP2}

\begin{figure}
\includegraphics*[scale=0.55,angle=0]{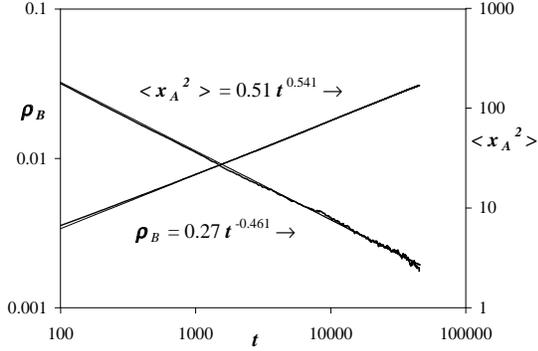}
\caption{\label{fig:dpc2} The $B$ particle density and $A$ species mean-square 
  displacement for the DP reactions $B \to \emptyset$ ($\lambda = 0.01$),
  $B \to 2 \, B$ ($\sigma = 0.2233$), $2 \, B \to \emptyset$ ($\mu = 1$) in
  $d = 2$.}
\end{figure}

\begin{figure}[b]
\includegraphics*[scale=0.5,angle=0]{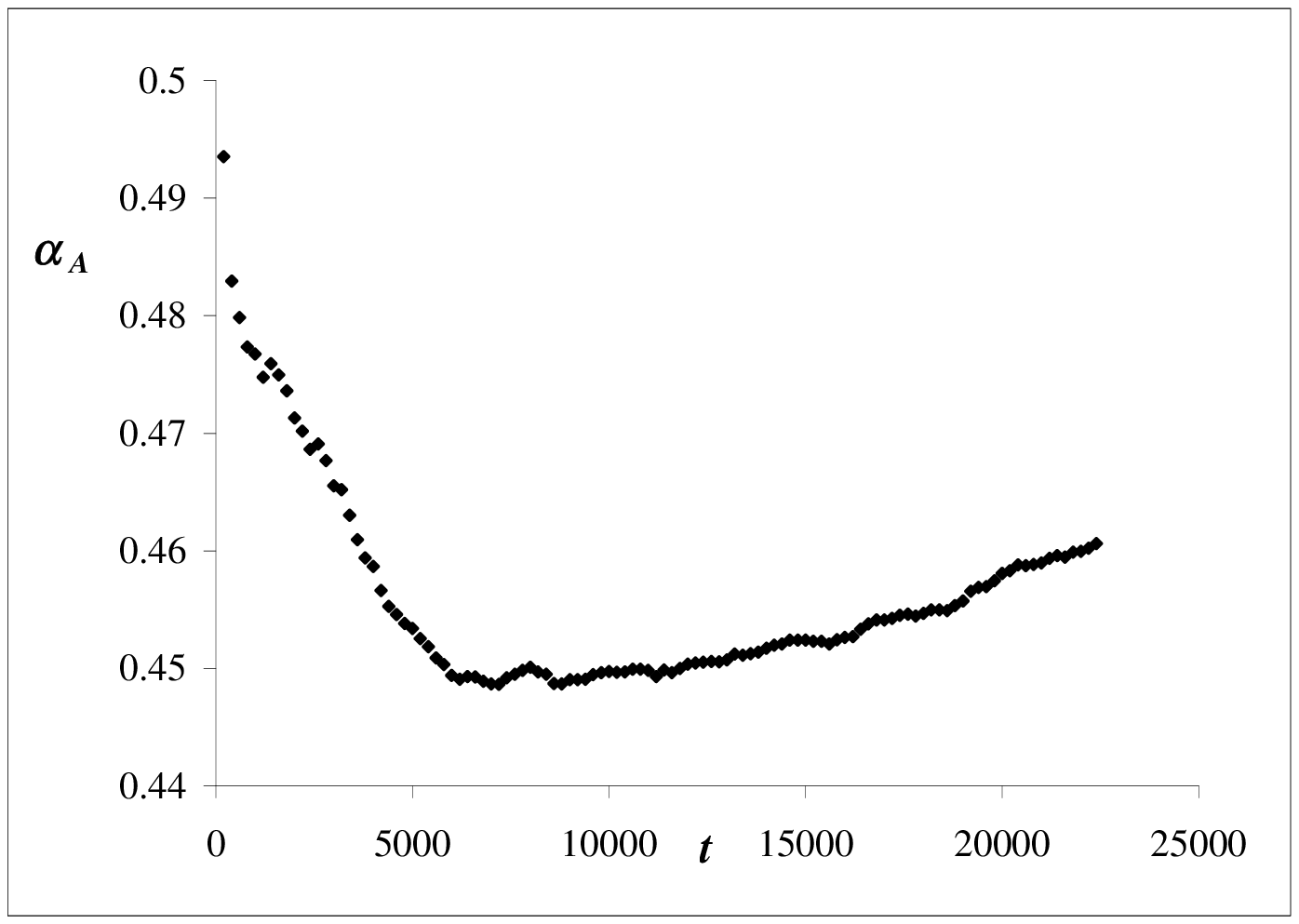}
\caption{\label{fig:dpc2a} The local exponent $\alpha_A$ corresponding to data
 shown in Fig.~\ref{fig:dpc2} (with $b=2$).}
\end{figure}

\begin{figure}
\includegraphics*[scale=0.5,angle=0]{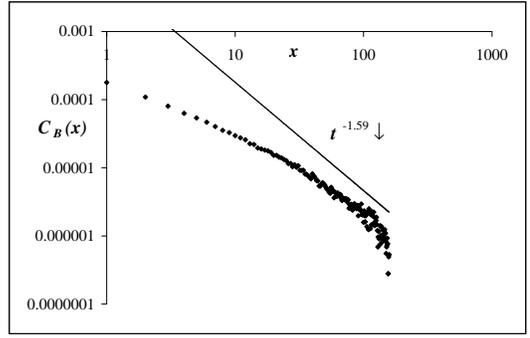}
\caption{\label{fig:dpf2} The $B$ particle pair correlation function $C_B(x)$
  at the DP critical point ($d = 2$, measured at $t = 50000$).}
\end{figure}

\begin{figure}[b]
\includegraphics*[scale=0.57,angle=0]{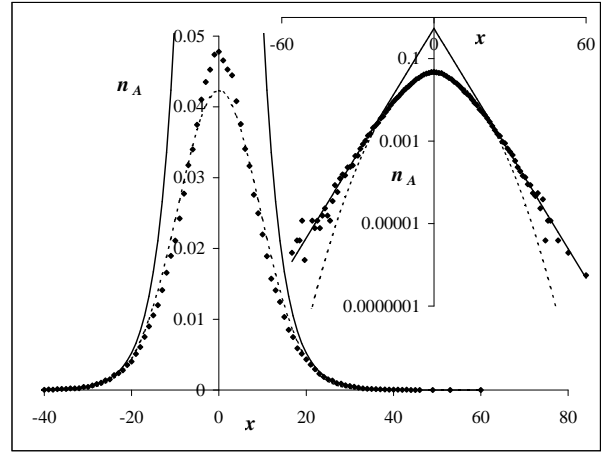}
\caption{\label{fig:dpd2} The measured distribution of the $A$ particle
  displacements corresponding to the system shown in Fig.~\ref{fig:dpc2} (DP 
  critical point for the $B$ system in $d = 2$, measured at $t = 50000$).
  The dashed line indicates a Gaussian fit, and the solid line is represents an
  unnormalized exponential fit to the distribution tails.}
\end{figure}

To investigate the properties at a DP transition in two dimensions, we used the
same reactions and rates as in $d=1$, namely $B \to \emptyset$
($\lambda = 0.01$), $B \to 2 \, B$ (varying $\sigma$), and
$2 \, B \to \emptyset$ ($\mu = 1$).
For $d = 2$, Tab.~\ref{exptb} predicts $\rho_B(t) \sim t^{-0.451}$, which
implies, according to our mean-field description, that
$\langle {\vec x}(t)_A^2 \rangle \sim t^{0.549}$.
Our best estimate of the critical branching rate is $\sigma_c \approx 0.2233$.
Figure~\ref{fig:dpc2} shows the power law dependence of $\rho_B(t)$ and
$\langle {\vec x}(t)_A^2 \rangle$, inferred from $40$ runs on a
$800 \times 800$ square lattice.
The initial $A$ density was set to $0.01$ (as opposed to the usual $0.5$) here.
The power law fits imply ${\widetilde D} = 0.256$, very close to the expected
$0.25$ from ordinary diffusion in two dimensions (since on average one $A$
particle will hop to each available $B$ site).
The slight error in computing ${\widetilde D}$ is probably a result of the
transient behavior at the beginning of the simulation, where the $B$ density is
still larger than the power law fit would predict.
Figure~\ref{fig:dpc2a} shows the local exponent $\alpha_A$ (computed with
$b=2$), indicating that the system is indeed subcritical, as $\alpha_A$ is
increasing with time for large $t$.
But the minimum plateau value is very close to the predicted
$\alpha_A \approx 0.45$.
Figure~\ref{fig:dpf2} demonstrates that the $B$ particle distribution is still
fractal at $t = 50000$, yielding an effective exponent value $2 \beta / \nu$
close to $1.59$ (as expected from DP) at intermediate distances.
Figure~\ref{fig:dpd2} depicts the $A$ species displacement distribution at
$t = 50000$ together with the mean-field Gaussian and an (unnormalized)
exponential fit of the distribution tails.
As in previous cases, the `tails' of the distribution are longer than a
Gaussian would suggest, and instead appear to be matched well by an
exponential.

At last, we verify the DP transition in two dimensions when $\lambda = 0$, and
compare the $A$ particle anomalous diffusion to the nonzero $\lambda$ case,
using the same reaction rates as before in $d = 1$.
Because $\alpha_B$ is so large for DP in $d=2$, we had to use quite sizeable 
systems ($800 \times 800$) to measure exponents over several decades in search 
of the critical point.
Thus our determination of $\sigma_c$ was not as precise.
For the $B$ particle density and $A$ species mean-square displacement as
functions of time, the measured exponents are $\alpha_B \approx 0.47$,
approximately $0.02$ larger than the DP value, and $\alpha_A \approx 0.48$.
We found our system close to but below criticality, and therefore observed
time-dependent local exponents $\alpha_A$ and $\alpha_B$.
We also completed fewer (20) runs for this system, and so expect a larger error
on our measurements of the exponents.
The power law fit prefactors imply ${\widetilde D} \approx 0.29$, in fair
agreement with our expectation of $0.25$ and previous measurements.
In summary, we have not uncovered any significant deviations from the behavior
observed for $\lambda > 0$.

\subsection{Parity conserving universality class, $d = 1$}
\label{PC}

\begin{figure}
\includegraphics*[scale=0.55,angle=0]{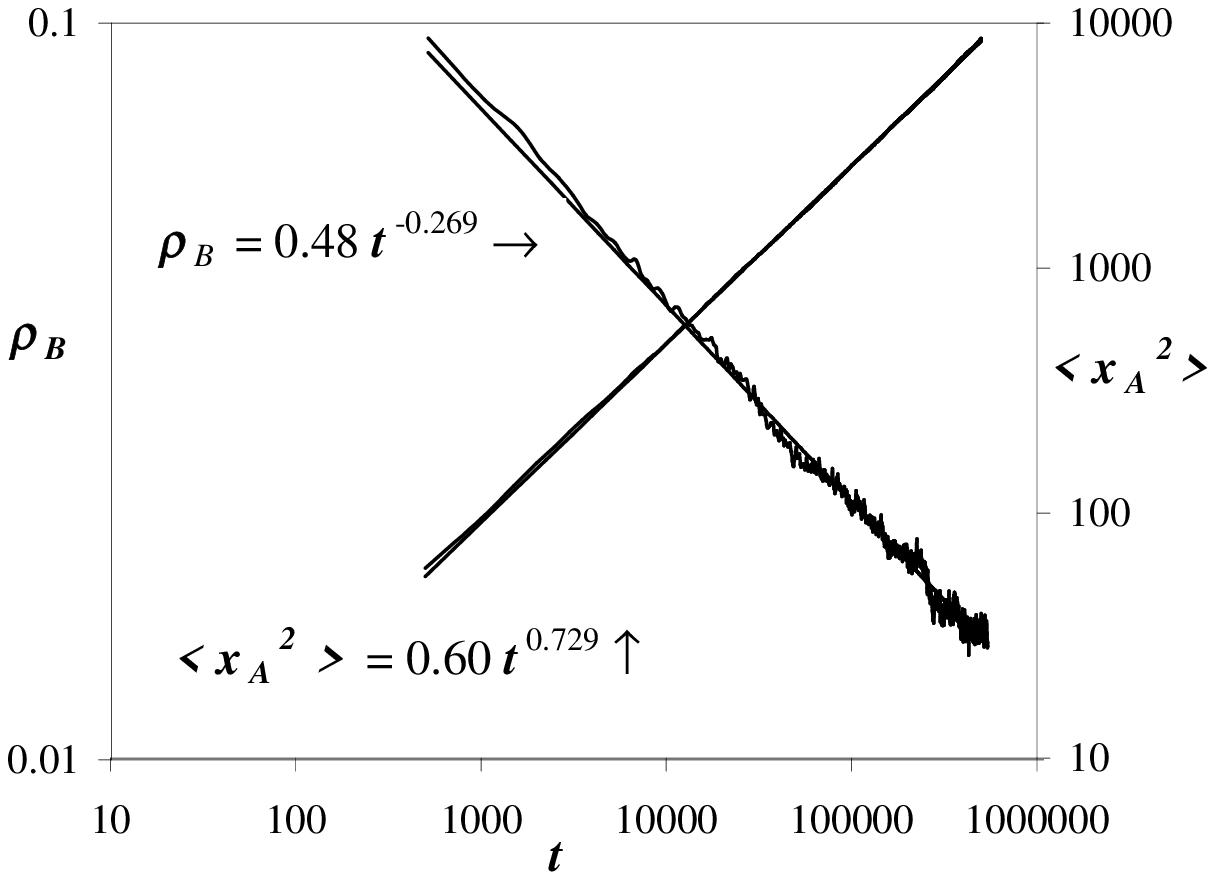}
\caption{\label{fig:pcc1} The $B$ particle density and $A$ species
  mean-square displacement for the PC reactions $B \to 3 \, B$
  ($\sigma = 0.2175$), $2 \, B \to \emptyset$ ($\mu = 0.5$) in $d = 1$.}
\end{figure}

\begin{figure}[b]
\includegraphics*[scale=0.5,angle=0]{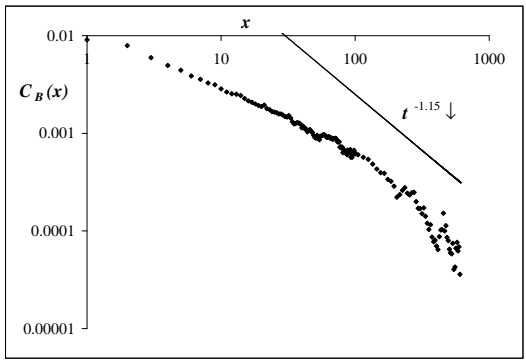}
\caption{\label{fig:pcc1a} The $B$ particle pair correlation function $C_B(x)$
at the PC critical point in $d = 1$ (measured at $t = 500000$).}
\end{figure}

In order to observe the phase transitions from active to inactive/absorbing
states in the PC universality class, we looked at branching and annihilating
random walks with even number of offspring particles, i.e., set $\lambda = 0$,
$n = 2$, and either $m = 2$ or $4$ in the reaction scheme (\ref{rxs}).
In the former case, $2 \, B \to \emptyset$ combined with $B \to 3 \, B$, we
were forced to set the annihilation probability to $\mu = 0.5$ in order to
detect the phase transition (at varying branching probability $\sigma$).
Figure~\ref{fig:pcc1} shows the power laws for the $B$ particle density decay
and the $A$ species mean-square displacement as a function of time near the
phase transition (at $\sigma_c \approx 0.2175$).
Our measurement of $\alpha_B = 0.269$ agrees fairly well with the expected PC 
exponent $0.285$ (from $20$ runs, each with $60000$ sites).
Furthermore, $\alpha_A = 0.271$ is in excellent agreement with $\alpha_B$.
The power law fits shown imply ${\widetilde D} = 0.45$ for this reaction.
Figure~\ref{fig:pcc1a} depicts a measurement of the $B$ density correlation
function at $t = 500000$, and the predicted PC exponent ratio $1.15$ according 
to Tab.~\ref{exptb}.
Measurements of $C_B(x)$ at $t = 50000$ and $t = 100000$ yielded distributions
very similar to that shown in Fig.~\ref{fig:pcc1a}.
Figure~\ref{fig:clus} in Sec.~\ref{intro} shows a space-time plot for these
reactions illustrating the fractal nature of the process at criticality.

\begin{figure}
\includegraphics*[scale=0.55,angle=0]{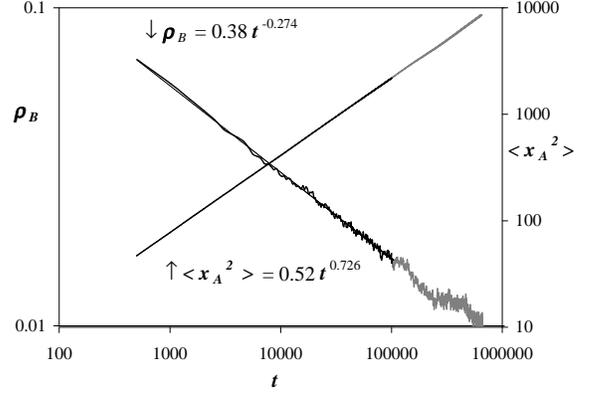}
\caption{\label{fig:pcc2} The $B$ particle density and $A$ species mean-square 
  displacement for the PC reactions $B \to 5 \, B$ ($\sigma = 0.2795$),
  $2 \, B \to \emptyset$ ($\mu = 1$) in $d = 1$.}
\end{figure}

\begin{figure}[b]
\includegraphics*[scale=0.57,angle=0]{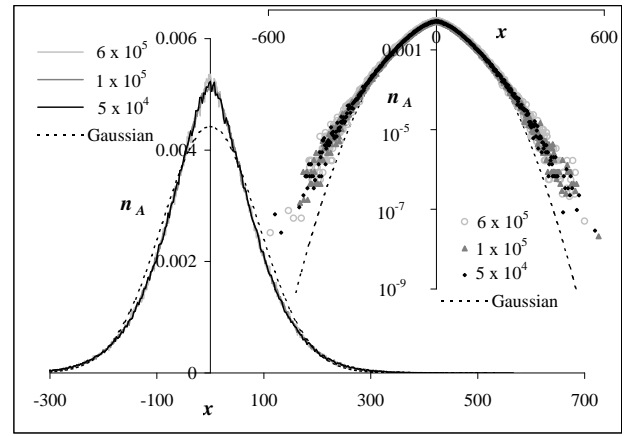}
\caption{\label{fig:pcd2} The measured distribution of the $A$ particle
  displacements corresponding to the system shown in Fig.~\ref{fig:pcc2} (PC 
  critical point for the $B$ system, $d = 1$) at different times as indicated.}
\end{figure}

\begin{figure}
\includegraphics*[scale=0.55,angle=0]{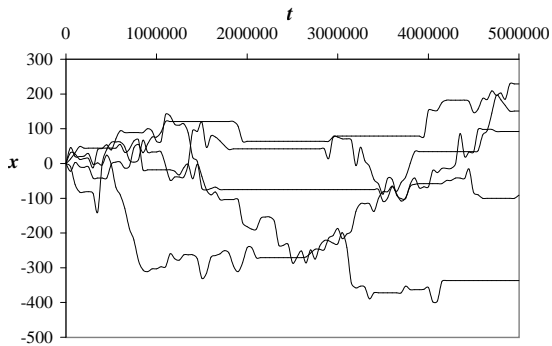}
\caption{\label{fig:pcst} The paths of the five $A$ particles with largest
  displacements in a simulation contributing to the data plotted in
  Fig.~\ref{fig:pcc2}.
  Notice the long intervals of particle localization interspersed with brief
  periods of high mobility.}
\end{figure}

\begin{figure}
\includegraphics*[scale=0.54,angle=0]{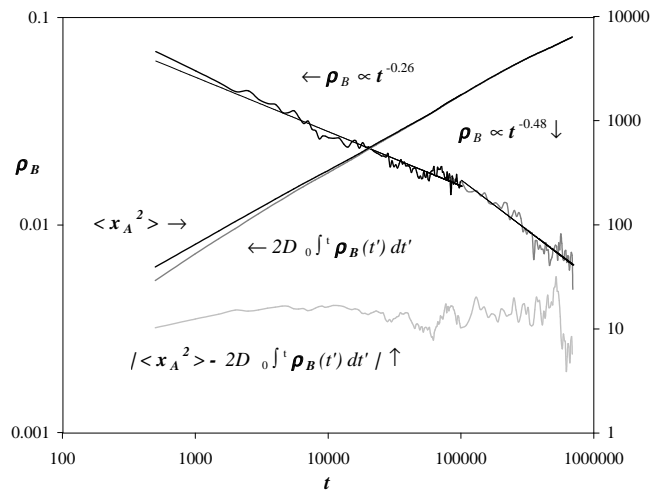}
\caption{\label{fig:pcs1} The $B$ particle density and $A$ species
  mean-square displacement for the PC system of Fig.~\ref{fig:pcc2} in the
  subcritical regime ($\sigma = 0.277 < \sigma_c$).
  The crossover from the critical power laws to those of the inactive phase is
  clearly visible.}
\end{figure}

We also observed the critical behavior for the PC reactions
$2 \, B \to \emptyset$ ($\mu = 1$) combined with $B \to 5 \, B$.
Setting $\mu = 1$ helps to minimize the number of necessary random numbers, as
well as prohibits the multiple occupation of lattice sites and thus avoids
discrepancies between the $B$ particle density and the density of available
sites for the $A$ species.
In Fig.~\ref{fig:pcc2} we display the $B$ density and $A$ species mean-square
displacement as a function of time near the phase transition (at
$\sigma_c \approx 0.2795$), implying ${\widetilde D} = 0.50$ (averaged over
$20$ runs on the lattice with $60000$ sites).
Indeed we again find good agreement $\alpha_B \approx 0.274 \approx \alpha_A$, 
as well as with the simulation data depicted in Fig.~\ref{fig:pcc1}, and the 
predicted PC value.
In addition, we measured the $A$ displacement distribution at three distinct
times: $t = 50000$, $t = 100000$, and $t = 600000$.
By scaling the ordinate by a factor $s = (t_2/t_1)^{(1-\alpha_A)/2}$, and the
abscissa by $1/s$, the three curves are seen to be essentially identical in
shape, see Fig.~\ref{fig:pcd2}, which supports the dynamic scaling conjecture
that the $A$ particle distribution maintains its shape as it evolves in time.
In Fig.~\ref{fig:pcst} we show the paths of five $A$ particles in one of the
simulations contributing to the data in Fig.~\ref{fig:pcc2}.
The selected $A$ particles are those with the greatest displacements.
The trajectories consist of periods of localization (very little activity)
interspersed with periods of much movement.
This diagram resembles those for other cases of anomalous diffusion such as
Levy flights.
We also examined a subcritical system, setting $\sigma = 0.277$.
Figure~\ref{fig:pcs1} shows an initial critical regime with 
$\rho_B(t) \sim t^{-0.26}$, crossing over very quickly to the subcritical
behavior dominated by pair annihilation, $\rho_B(t) \sim t^{-0.5}$.
We see that throughout these regimes the mean-square displacement of the $A$
particles agrees remarkably well with the integral of the $B$ density, setting
${\widetilde D} = 0.425$ (from $7$ runs, each with $40000$ sites).

\begin{figure}
\includegraphics*[scale=0.6,angle=0]{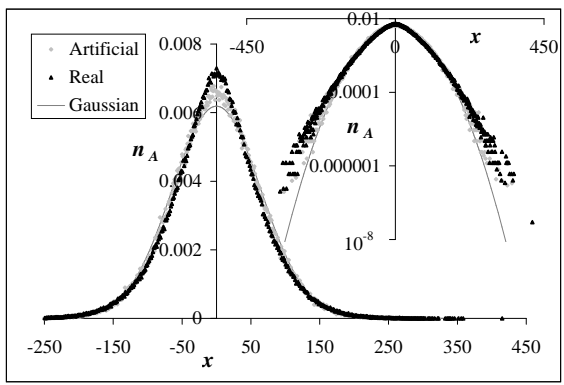}
\caption{\label{fig:pcar} The measured distribution of the $A$ particle
  displacements corresponding to the system shown in Fig.~\ref{fig:pcc2}
  (PC critical point for the $B$ system, $d = 1$), compared with the one
  resulting from the artificial model with uncorrelated $B$ particles.
  Notice that different effective diffusivities were applied to obtain equal
  second moments.}
\end{figure}

To uncover the origins of the non-Gaussian distributions observed in the
systems thus far examined, we also considered a simple model in which the $B$
particle distribution remained uncorrelated, while still exhibiting the desired
overall density decay.
The algorithm allowed the $B$ species to diffuse through the lattice without
site exclusion, but removed sufficiently many of them at random so that the $B$
density followed the prescribed global behavior.
For example, we forced the $B$ density to decay as observed in
Fig.~\ref{fig:pcc2}, near the PC phase transition in $d=1$.
As would be expected, the time dependence of the mean-square displacement of
the $A$ species agreed with the integral of the $B$ particle density with the
absolute error bounded by $10$.
The required choice for this agreement was ${\widetilde D} = 0.95$, a
puzzlingly large value, considering all previous observations gave
${\widetilde D \approx 0.5}$.
Since in this algorithm the $B$ particles are not being continuously created
and annihilated, the $A$-$B$ coupling is actually stronger, and thus on average
there is probably a higher density of $A$ particles surrounding a typical $B$
site, thereby increasing the effective $A$ diffusivity.
We may now compare the $A$ displacement distributions at $t = 100000$ for the
homogeneous and true PC systems, see Fig.~\ref{fig:pcar}.
Different effective diffusivities required an area-preserving rescaling.
While there does appear to be some deviation from a Gaussian even in the
artificial model, the figures suggest that the deviation is not as great as in
the true PC system with $B$ particle correlations.
Though not shown, a rescaled version of the radioactive decay $A$ displacement
distribution agrees well with the `artificial' distribution shown here, as one
might expect.
We suggest that the low connectivity of the one-dimensional lattice necessarily
causes significant deviations from the average effective diffusivity
${\widetilde D} \langle \rho_B(t) \rangle$ for the $A$ species, which then
yields a non-Gaussian distribution of the $A$ particle displacements.
However, Fig.~\ref{fig:pcar} demonstrates that $B$ particle correlations are
also significant in producing deviations from the mean-field prediction.

We also explored a few other algorithms to verify that our results were not
particular to our implementation.
First we adapted the $A$ species hopping algorithm, so that whenever an $A$
particle could hop, it would with probability $1$.
Previously, $A$ particle hopping was implemented by choosing a direction to
hop, and then checking to see if the destination nearest neighboring site was
occupied by a $B$ particle.
We executed the maximal $A$ particle hopping algorithm for $\sigma = 0.2795$,
our best estimate of the critical point for the PC system with $m=4$.
The time dependence of the mean particle displacement agreed very closely with
the integral of the $B$ particle density, with ${\widetilde D} = 0.87$.
At $t = 100000$ the $A$ displacement distribution coincided with that in
Fig.~\ref{fig:pcd2}, apart from a scaling factor to account for the discrepancy
in ${\widetilde D}$.
The increase in this value simply indicates that $A$ particles will hop 
whenever a neighboring site is available.

Next we evolved the $B$ particles until $t = 500$, when the $B$ density was
down to $0.07$.
Thus clusters and relatively vacant regions should have begun to form.
The algorithm then placed $A$ particles on the existing $B$ particles and
studied the $A$ diffusion thereafter.
Though after a longer transient, the mean-square $A$ displacement approached
the integral of the $B$ density (setting ${\widetilde D} = 0.5$).
Finally we examined an algorithm that removed $A$ particles from the system
after they had been localized for a certain time interval (here chosen as
$10000$ Monte Carlo steps).
We found the number of active $A$ particles to be a sharply decreasing function
of time.
The localized particles being removed from the computation had lower effective
diffusion rates than the remaining active particles, overall increasing the
mean-square $A$ displacement.
This observation supports the hypothesis that the effective diffusion rate for
the $A$ species is not uniform, indicating that in the previous models, a
significant fraction of the $A$ species was localized over time scales at least
as large as 10000 Monte Carlo steps.
Thus most $A$ particles are not truly diffusing through the dynamic fractal,
but rather execute an occasional hop when passed over by $B$ particles.

\section{Concluding remarks}
\label{concl}

\begin{figure}
\includegraphics*[scale=0.55,angle=0]{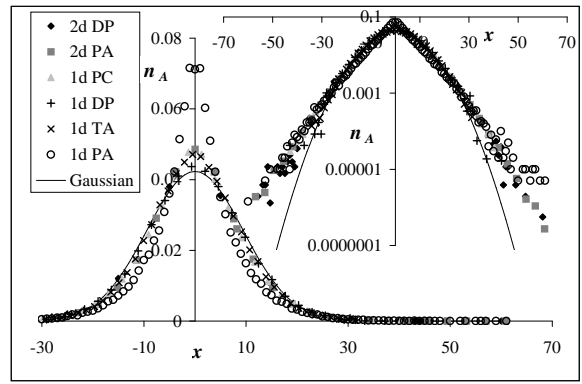}
\caption{\label{fig:difd} The measured $A$ species displacement distributions
  for the majority of systems investigated here (PA and TA represent the pure
  $B$ particle pair and triplet annihilation processes, respectively).
  We see that most of the data collapse to roughly the same scaling function
  (except for the result for one-dimensional $B$ pair annihilation with the
  strongest anti-correlations).
  The observed distribution clearly deviates from the mean-field Gaussian both
  at small and large displacements.}
\end{figure}

We have numerically investigated the reaction-controlled diffusion model
introduced in Ref.~\cite{trimper00} for various $B$ species reaction systems,
and studied the ensuing anomalous diffusion for the $A$ particles on the
emerging dynamic fractal $B$ clusters.
We found excellent agreement for all examined systems of the $A$ species 
mean-square displacement $\langle {\vec x}(t)^2_A \rangle$ with the integral of
the mean B particle density $\langle \rho_B(t) \rangle$, at least within the
accuracy of our data.
However, the mean-field Gaussian distribution (\ref{sol}) for the $A$ particle
displacements was not observed.
We have argued that the primary factor in creating this deviation is a
variation in the effective diffusion rate of the $A$ particles, which is
proportional to the number of hops executed by an $A$ throughout the
simulation run.
The shape of the resulting $A$ displacement distribution was seen to be the
same at different times and also over many different systems (see
Fig.~\ref{fig:difd}), thus suggesting that the distribution shape of effective
diffusion rates also remains the same over different time scales and systems.

\begin{figure}
\includegraphics*[scale=0.55,angle=0]{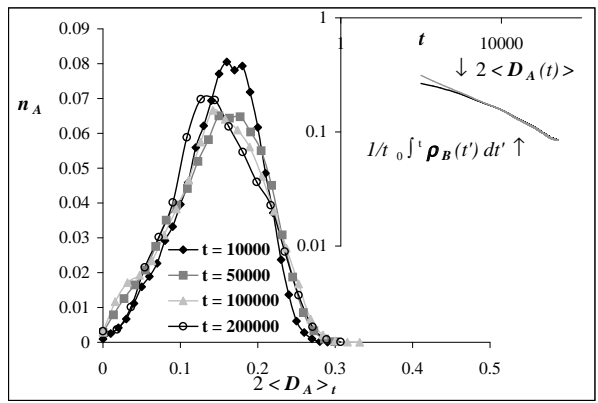}
\caption{\label{fig:difr} The distribution of effective $A$ diffusivities,
  obtained as the average number of hops per unit time, at various times during
  simulations for critical DP processes for the $B$ system in $d = 1$.
  The inset shows that the average diffusion coefficient becomes identical with
  the temporal average of the $B$ species density.
  The diffusion rate distributions are rescaled by to the same average for
  better comparison.}
\end{figure}

The contributing factors to producing the diffusion rate distribution were
identified as persisting spatial fluctuations in the local $B$ particle density
(enhanced in situations with strong $B$ correlations), the low connectivity of
the one- and two-dimensional lattices examined here, and the fact that the $A$
species tend to be carried along with diffusing $B$ particles (the `piggy-back'
effect).
As mentioned above, the distribution of effective diffusivities is constrained
in that the average diffusion rate should equal
${\widetilde D} \, \langle \rho_B(t) \rangle$.
The evolution of this average diffusion rate according to
$t^{-1} \int_{0}^{t} \rho_B(t') dt'$ (see Fig.~\ref{fig:difr}) coupled with the
time independence of the $A$ distribution shape, i.e., dynamic scaling, 
essentially yields the mean-field time behavior of 
$\langle {\vec x}^2_A(t) \rangle$, Eq.~(\ref{m2s}).
Also, as the mean-square displacement is an integral quantity, the noise
associated with fluctuations in the $B$ particle density becomes suppressed.
Consequently, under the assumption that $\alpha_A = \alpha_B$ to a very good
approximation at least, one may measure critical exponents of the reacting $B$
particle system via the passive $A$ species using fewer simulation runs and 
with reduced statistical noise.

Fluctuation effects really become manifest in the higher moments only, and in
the overall shape of the $A$ displacement distribution.
Figure~\ref{fig:difd} shows the $A$ species displacement distributions
$n_A({\vec x},t)$ measured at different times from most of the systems examined
thus far (apart from the case of radioactive $B$ decay), all scaled to have
approximately the same second moment $\langle {\vec x}^2_A \rangle$.
Aside from the case of one-dimensional $B$ pair annihilation with its strong
anticorrelations, all distributions appear to at least approximately collapse
to the same scaling function.
As we have seen previously, compared with a Gaussian this common distribution 
features an excess of particles both in its `tails' and peak.
For the one-dimensional pair annihilation case these deviations are markedly
enhanced.
We find this apparent universality in the $A$ particle displacement
distribution quite surprising.
Yet in each case depicted in Fig.~\ref{fig:difd}, correlations developed
between the $B$ particles, leading to significant inhomogeneities in their
spatial distribution.

However, assuming that the $A$-$B$ coupling is sufficiently weak that
approximate $A$ spatial homogeneity is maintained, the distribution of the
time-averaged effective $A$ diffusivities is constrained by the global value
${\widetilde D} \, \langle \rho_B(t) \rangle$.
We compared these quantities for a one-dimensional DP system (system size
10000, 10 runs with $\lambda = 0.01$) and indeed saw excellent agreement at
long times (see the inset of Fig.~\ref{fig:difr}), indicating that at each time
step on average one $A$ particle hops to each $B$ site.
The discrepancy at small times can be attributed to numerical errors in 
computing the integral.
Apparently the time-averaged diffusion rate distribution over the $A$ particles
is common to the bulk of the reaction systems studied here.
To address the apparent universality in systems of positive or weak (but
existing) $B$ species correlations, we should note that localization, by which
we mean the annihilation of the nearest $B$ particle to the newly localized
$A$, is often impermanent when the $A$ resides in a $B$ cluster that results
from positive correlations.
Figure~\ref{fig:difr} shows the effective $A$ diffusion rate distribution at
various times for the one-dimensional DP system.
The distribution at $t = 10000$ is still fairly peaked, which may be accounted
for by the initial high density of $B$ particles leading to homogenization of
the diffusion rates.
The distributions at later times (scaled to match the average value of the
$t = 10000$ distribution) seems approximately constant, though perhaps the peak
is slowly shifting towards smaller values.
Finally, we note that the measured values of ${\widetilde D}$ were 
approximately the same for all systems, namely ${\widetilde D} \approx 0.5$ in 
$d = 1$ and ${\widetilde D} \approx 0.25$ in $d = 2$.
\vfill

\begin{acknowledgments}
This research has been supported through the National Science Foundation,
Division of Materials Research, grant no. DMR 0075725, and the Jeffress 
Memorial Trust, grant no. J-594.
We gladly acknowledge helpful discussions with Olivier Deloubri\`ere, 
Jay Mettetal, Beate Schmittmann, and Royce Zia.
\end{acknowledgments}


\begin{thebibliography}{}
\bibitem{fnote3}
  Since we allow any number of B particles to occupy the same lattice site, 
  the last reaction in (\ref{rxs}) is required to maintain a finite particle 
  density in the active phase. Alternatively one could apply a restriction of 
  the site occupation numbers, e.g., to $0$ and $1$ only.
\bibitem{fnote2}
   Henceforth, $\rho_A$ and $\rho_B$ denote the $A$ and $B$ particle number
   densities, and we shall refer to $\lambda$, $\sigma$, and $\mu$ as the
   probabilities (rather than rates) for spontaneous $B$ particle decay,
   offspring production, and $n$th order annihilation processes, respectively.
\end{thebibliography}
\end{document}